\documentclass[sigconf]{acmart}

\AtBeginDocument{%
  }

\copyrightyear{2026}
\acmYear{2026}
\setcopyright{cc}
\setcctype{by}
\acmConference[CHI '26]{Proceedings of the 2026 CHI Conference on Human Factors in Computing Systems}{April 13--17, 2026}{Barcelona, Spain}
\acmBooktitle{Proceedings of the 2026 CHI Conference on Human Factors in Computing Systems (CHI '26), April 13--17, 2026, Barcelona, Spain}
\acmPrice{}
\acmDOI{10.1145/3772318.3791724}
\acmISBN{979-8-4007-2278-3/2026/04}

\usepackage{algorithm}
\usepackage{algorithmic}

\definecolor{zzhpurple}{RGB}{128,0,128}

\newcommand{\name}{\textsc{DiagLink}}
\newcommand{\term}[1]{\textit{\textsc{#1}}}

\newcommand{\zzh}[1]{\textcolor{zzhpurple}{#1}}

\definecolor{myblue}{RGB}{0,102,204}

\newcommand{\rr}[1]{{\color{myblue}#1}}

\newcommand{\rrfig}[2][]{%
  \begingroup
  \setlength{\fboxrule}{1pt}
  \setlength{\fboxsep}{0pt}
  \fcolorbox{myblue}{white}{\includegraphics[#1]{#2}}%
  \endgroup
}

\newcommand{\rrtab}[1]{%
  \begingroup
  \setlength{\fboxrule}{1pt}
  \setlength{\fboxsep}{2pt}
  \fcolorbox{myblue}{white}{#1}%
  \endgroup
}

\renewcommand{\rr}[1]{#1}                            
\renewcommand{\rrfig}[2][]{\includegraphics[#1]{#2}} 
\renewcommand{\rrtab}[1]{#1}                         
\renewcommand{\zzh}[1]{#1}

\begin{document}

\title{\name: A Dual-User Diagnostic Assistance System by Synergizing Experts with LLMs and Knowledge Graphs}

\author{Zihan Zhou}
\authornote{Both authors contributed equally to this research.}
\email{zhouzihan.liu@gmail.com}
\orcid{0009-0006-5861-7689}
\author{Yinan Liu}
\authornotemark[1]
\email{liuyinan@cse.neu.edu.cn}
\orcid{0009-0002-3316-3693}
\affiliation{%
  \institution{Northeastern University}
  \department{School of Computer Science and Engineering}
  \city{Shenyang 110819}
  \country{China}
}

\author{Yuyang Xie}
\email{meguriri1019@gmail.com}
\orcid{0009-0006-8063-0388}
\affiliation{%
  \institution{Northeastern University}
  \department{School of Computer Science and Engineering}
  \city{Shenyang 110819}
  \country{China}
}

\author{Bin Wang}
\email{binwang@mail.neu.edu.cn}
\orcid{0000-0002-2694-1023}
\affiliation{%
  \institution{Northeastern University}
  \department{School of Computer Science and Engineering}
  \city{Shenyang 110819}
  \country{China}
}

\author{Xiaochun Yang}
\email{yangxc@mail.neu.edu.cn}
\orcid{0000-0002-6184-4771}
\affiliation{%
  \institution{Northeastern University}
  \department{School of Computer Science and Engineering}
  \city{Shenyang 110819}
  \country{China}
}

\author{Zezheng Feng}
\email{fengzezheng@swc.neu.edu.cn}
\orcid{0000-0003-4874-8133}
\authornote{Corresponding author.}
\affiliation{%
  \institution{Northeastern University}
  \department{Software College}
  \city{Shenyang 110819}
  \country{China}
}

\newcommand{\eg}{\emph{e.g.}}
\newcommand{\ie}{\emph{i.e.}}
\newcommand{\sect}{Sect.~}
\newcommand{\fig}{Fig.~}
\newcommand{\eq}{Eq.~}

\newcommand{\fzz}[1]{{\color{red}{#1}}}
\renewcommand{\shortauthors}{Zihan Zhou et al.}

\begin{abstract}

The global shortage and uneven distribution of medical expertise continue to hinder equitable access to accurate diagnostic care. 
While existing intelligent diagnostic system have shown promise, most struggle with dual-user interaction, and dynamic knowledge integration—limiting their real-world applicability. 
In this study, we present \name, a dual-user diagnostic assistance system that synergizes large language models (LLMs), knowledge graphs (KGs), and medical experts to support both patients and physicians. 
\name~uses guided dialogues to elicit patient histories, leverages LLMs and KGs for collaborative reasoning, and incorporates physician oversight for continuous knowledge validation and evolution. 
The system provides a role-adaptive interface, 
dynamically visualized history, 
and unified multi-source evidence to improve both trust and usability.
We evaluate \name~through \rr{user study,} use cases and expert interviews, demonstrating its effectiveness in improving user satisfaction and diagnostic efficiency, while offering insights for the design of future AI-assisted diagnostic systems.

\end{abstract}





\maketitle

\section{Introduction}

\label{intro}

The scarcity and unequal distribution of high-quality diagnostic resources have long posed a huge challenge for healthcare systems worldwide~\cite{Fleming2021Diagnostics, Wilson2018PALM, Hricak2021Imaging}. 
By 2022, the global average physician density was 17.2 per 10,000 people, with more than 40\% of countries falling below 10 per 10,000~\cite{WHO2025PhysicianDensity}. In 2023, the global shortage of healthcare workers remained as high as 14.7 million~\cite{WHO2024EB156}. In rural and remote regions, up to two-thirds of the population lack access to essential services~\cite{WHO2021RuralGuideline}, leading to delays in treating conditions that could often be mitigated through early intervention~\cite{Symington2025}. Moreover, marked disparities in diagnostic skills and quality further compound these risks, with an estimated $795,000$ cases of serious harm or death attributable to diagnostic errors reported annually in the United States alone~\cite{NewmanToker2024}.

To address this challenge, developing and leveraging intelligent diagnostic systems to improve diagnostic efficiency, optimize workforce allocation, and narrow competency gaps has emerged as a promising solution. Early studies have explored clinical decision support systems (CDSSs) based on knowledge graphs (KGs), which have been employed in disease diagnosis and medication recommendation~\cite{Nelson2022SPOKEsig, Xu2023SeqCare, Tang2024NatAging}.
By structuring medical concepts (\eg, diseases, symptoms, and medications) and modeling their semantic relationships, KGs can provide explainable reasoning support for clinical decision-making~\cite{Chandak2023PrimeKG}. Despite these strengths, KGs are inherently constrained in processing unstructured clinical text and capturing complex contextual semantics. 
Recently, large language models (LLMs), with their powerful natural language processing capabilities, have shown remarkable advantages across tasks such as medical history collection~\cite{tu2025towards}, clinical question answering~\cite{zhao2025medrag}, and biomedical literature analysis~\cite{Oami2024JAMANetwOpen}. However, LLMs lack explainability and factual consistency~\cite{sreekar2024axcel,luo2025etrqa}. Reasoning results generated by LLMs can often not be verified through LLMs' internal parametric knowledge. Intuitively, the complementary potential between LLMs and KGs provides a compelling rationale for hybrid diagnostic systems that combine the contextual language understanding of LLMs with the explicit, structured knowledge offered by KGs.

In AI research community, previous studies have explored integrating the structured knowledge from KGs into language models to improve the quality of generated outputs. Existing approaches can be categorized into three directions: (1) retrieving relevant entities and relations from KGs to enhance the accuracy and consistency of generated content~\cite{wen2023mindmap, wu2025medreason}; 
(2) injecting structured constraints during the decoding stage to reduce improper inferences~\cite{luo2024graph}; and 
(3) employing LLMs as agents to explore KGs for improving reasoning completeness~\cite{sun2023think, ma2024think}.
Concurrently, the visualization and human-computer interaction community has sought to combine the conversational strength of LLMs with the structured characteristics of KGs to improve knowledge presentation and verification.
Previous studies have explored linking dialogue systems with KGs to facilitate information retrieval~\cite{yan2024knownet, li2024linkq}, utilizing KG-based visualization and interactive exploration to realize personalized recommendation~\cite{gao2025healthgenie}, and incorporating multi-agent mechanisms with KGs to improve the reliability of query responses~\cite{li2025accurate}.
While these methods show promising results in offline evaluations or task-specific settings, they remain insufficient for real-world diagnostic support due to several limitations: 
(1) they are often designed for isolated subtasks (\eg, retrieval or reasoning) and lack end-to-end support for joint patient-physician use; 
(2) they typically rely on static KGs that cannot dynamically incorporate evolving medical knowledge; and 
(3) they operate without continuous expert input, limiting the reliability and clinical relevance of diagnostic suggestions.
It is worth noting that the preliminary reasoning results generated through collaboration between the LLM and the KG can provide valuable support to expert diagnosis. At the same time, experts are able to verify and correct these results. Crucially, expert feedback along with new knowledge—such as newly discovered disease associations—can be used to update and expand the KG. In this way, the entire system continuously improves and evolves under the guidance of human experts, much like a snowball effect.
Therefore, a critical unresolved issue is the effective integration of LLMs, KGs, and medical experts into a collaborative framework to support both patients and physicians, which is an essential step toward alleviating the structural shortage of healthcare resources.

In this paper, we develop \name, a novel dual-user interactive diagnostic assistance system that connects patients and physicians through the synergy of LLMs, KGs, and experts. 
\rr{First, \name\ uses emotionally supportive guided dialogue to systematically elicit medical histories within a semi-structured template, easing patients’ communication burden and helping them make sense of their condition, while enabling physicians to focus more on key diagnostic decisions.}
During diagnosis, \name\ establishes a closed-loop reasoning workflow that leverages the complementary strengths of LLMs, KGs, and medical experts. The LLM generates initial diagnostic results based on its internal knowledge, and the KG expands diagnostic diseases through structured reasoning and subsequently refines them under the retrieval-augmented generation (RAG) paradigm. 
While experts further contribute to driving continuous knowledge evolution via subgraph generation, redundancy checks, and validation techniques. 
\name\ then presents patient information, \rr{predicted diagnoses}, and \rr{relevant evidence} through an integrated text–graph interface.
Using relevance- and importance-based node ranking combined with progressive exploration, \name~guides physicians toward critical insights, reduces cognitive load, and facilitates final diagnostic decisions.
\rr{Finally, we evaluate \name\ through a controlled user study, complemented by quantitative case analyses and expert interviews, demonstrating its ability to enhance patients’ diagnostic experience and improve physician efficiency, while also moderately boosting diagnostic accuracy.}

In summary, our main contributions are listed as follows:
\begin{itemize}
    \item We design and implement a novel dual-user interactive diagnostic assistance system serving both patients and physicians through the collaboration of LLMs, KGs, and experts.
    \item We propose a role-adaptive interface that supports perceptible medical history collection for patients and provides linked multi-source evidence for physicians’ diagnosis.
    \item We demonstrate the effectiveness of \name\ through \rr{user study,} use cases and expert interviews, further validating its potential to improve patient–physician communication, enhance diagnostic efficiency,  and inform the design of future collaborative medical AI systems.
\end{itemize}

\section{Related Work}

\subsection{Intelligent Diagnostic Support Systems}

Intelligent diagnostic support systems (IDSSs) are regarded as an important technological pathway to alleviating disparities in healthcare resources, providing physicians and patients with scalable diagnostic assistance through algorithms~\cite{Li2025Automatic}, KGs, and interactive visualization interfaces.
Existing IDSSs mainly consist of online symptom checkers (OSCs) for patients~\cite{wallace2022diagnostic} and clinical decision support systems (CDSSs) for physicians~\cite{musen2021clinical}, both of which have established reusable technical and interaction paradigms for improving diagnostic accessibility and efficiency.
For instance, Ada OSC ~\cite{AdaHealth2025} employs a conversational front-end integrated with an inference engine based on Bayesian networks and a medical KG, which adaptively follows up with patients to collect symptoms and signs, and generates probabilistically ranked candidate diagnoses and triage recommendations.
In evaluations on U.S. emergency department patient data, the system demonstrated a Top-3 diagnostic accuracy of approximately 63\%, a triage agreement rate of 62\%, and an unsafe triage rate of 14\%~\cite{fraser2023comparison}. In a rheumatology outpatient trial, the median completion time was about 7 minutes with relatively high usability ratings, though its diagnostic sensitivity was limited by the scope of disease coverage~\cite{knitza2021accuracy}.
In the domain of CDSS, the Targeted Real-Time Early Warning System (TREWS), a machine-learning–based sepsis early warning tool embedded in the EHR with real-time scoring and alert interfaces and integrated into clinical workflows, was associated with a relative reduction of 18.7\% in in-hospital mortality among sepsis patients in a multi-center real-world study~\cite{adams2022prospective}.
Overall, such systems have demonstrated clinical value in improving accessibility and efficiency, but diagnostic accuracy, explainability, and the design of safety guardrails  remain critical factors influencing trust and adoption~\cite{sutton2020overview, ouanes2024effectiveness}. Moreover, most existing studies focus on isolated components such as information retrieval, question answering, or reasoning, while lacking end-to-end interaction designs that encompass medical history taking, diagnostic recommendation generation, and evidence presentation.

\subsection{Construction and Applications of Medical KGs}

Medical KGs organize vast, fragmented medical knowledge into entity–relation schema that interconnect core concepts such as diseases, symptoms, and medications, thereby providing structured support for clinical decision-making~\cite{Chandak2023PrimeKG, Devarakonda2024CTKG}. 
The construction of medical KGs through automated extraction and representation from heterogeneous medical data has attracted extensive research attention.
Rotmensch et al.~\cite{rotmensch2017learning} proposed a method for learning KGs from electronic health records (EHRs), leveraging implicit associations in clinical notes to generate structured knowledge and demonstrating the potential of data-driven medical knowledge discovery.
Roy and Pan~\cite{roy2021incorporating} incorporated medical knowledge into BERT to enhance clinical relation extraction, improving the accuracy of entity–relation identification in clinical text. 
Building on large-scale pretrained language models, Harnoune et al.~\cite{harnoune2021bert} conducted knowledge extraction and analysis to construct a biomedical KG, presenting the potential of such models for KG construction and downstream analysis.

On the application side, KGs have been widely adopted to support knowledge exploration, clinical reasoning, and knowledge governance. 
Husain et al.~\cite{husain2021multi} proposed a multi-scale visual analytics framework, allowing researchers to switch between global graph layouts and local neighborhood details, while integrating literature evidence to uncover potential associations.
KGScope~\cite{yuan2024kgscope} lowers the barrier to exploration through embedding-guided navigation and interactive cues. 
Through interviews with frontline KG practitioners, Li et al.~\cite{li2023knowledge} summarized
 interaction and visualization~\cite{feng2022survey,feng2024holens} challenges across the “build–explore–analyze” pipeline and distilled transferable design principles for healthcare scenarios. 
Fan et al. ~\cite{fan2024visual} presented an interactive visual comparison method for multi-outcome causal graphs, helping physicians reason about cross-outcome inference paths under multimorbidity. 
Kou et al.~\cite{kou2022hc} proposed HC-COVID, a hierarchical crowdsourcing framework that builds a KG to detect and explain pandemic misinformation, underscoring the value of human–AI collaboration for knowledge updating and verification. 
Shang et al.~\cite{shang2024electronic} further integrated cross-institutional electronic health record data to develop a KG–driven decision support system, which presents patient-specific pathways and multi-source evidence through a visual interface, demonstrating the potential of KGs in real-world clinical workflows.

Overall, substantial progress has been made in knowledge visualization, interactive exploration, and evidence-chain construction. 
Nonetheless, the rapid evolution of medical knowledge challenges static KGs \cite{shen2018predicting,liu2020named}.
Existing update methods, typically dependent on structured data and predefined rules, struggle to incorporate novel medical knowledge~\cite{xin2024kartgps, hao2023hofd}, despite recent advances in dynamic~\cite{xu2025fast} and incremental learning~\cite{Ma2025Defying}..
Prior work~\cite{kou2022hc} has emphasized involving domain experts in knowledge updates, a key open challenge remains: how to seamlessly integrate expert knowledge into KG updating during use to enable continual \cite{liu2021joint}, trustworthy evolution.

\subsection{Integrating KGs with LLMs for Enhanced Usability}
Recently, LLMs have demonstrated remarkable capabilities in tasks such as medical history taking, clinical question answering, and biomedical literature analysis~\cite{fahrner2025generative}, driven by evolving transformer architectures~\cite{Ge2025Gated}.". 
However, their inherent black-box nature raises concerns regarding hallucination and explainability~\cite{huang2025survey, zhao2024explainability,feng2024trafps}. Consequently, recent studies  are focusing on integrating KGs as external knowledge supports and reasoning constraints for LLM outputs.
MedGraphRAG~\cite{wu2025medical} organizes biomedical literature, controlled terminologies, and user materials into a tripartite KG, using hierarchical retrieval with multi-hop evidence constraints to improve reliability and accuracy in medical QA and health fact-checking.
GCR~\cite{luo2024graph} integrates KG structures into the LLM decoding process via prefix-tree indexing to enforce graph-consistent reasoning. 
ToG~\cite{sun2023think} treats LLMs as agents that navigate and expand KG reasoning paths through beam search to strengthens reasoning completeness.

Meanwhile, the visualization and HCI communities have also begun exploring the integration of LLMs and KGs, leveraging the natural language capabilities of LLMs and the structured representations of KGs.
KnowNet~\cite{yan2024knownet} introduced a visual exploration system that integrates LLMs with medical KGs, enabling users to switch between natural language dialogue and graph-based evidence to support health information retrieval and validation.
LinkQ~\cite{li2024linkq} introduced an interactive interface for KG-based question answering that transforms LLM-generated queries into graph queries and validates them through visualization. 
HealthGenie~\cite{gao2025healthgenie} integrated a dietary KG with LLMs to create an interactive recommendation system that visualizes semantic links between ingredients and health goals to enhance user understanding and trust.
Zhang et al.~\cite{li2025accurate} designed a human–AI collaborative multi-agent system that combines KG-based evidence tracing with LLM-based natural language interaction, enhancing transparency and trust in multi-stakeholder medical decision-making.
\rr{Recent work on KG visualization and LLM–KG synergy further demonstrates interactive graph interfaces and causal KGs for talent management, mental health, and LLM knowledge organization~\cite{xu2025interactive,xu2025demo,meng2025deconstructing,yang2023visualization}.}

Overall, integrating LLMs with KGs improves diagnostic accuracy and strengthens user trust through visualization and interaction. 
Yet in high-stakes domains such as healthcare, models alone cannot ensure safety~\cite{morey2025empirically, sokol2025artificial}. Incorporating clinical experts enables a triadic collaboration—LLMs for language processing, KGs for structured reasoning, and experts for supervision—that balances efficiency, interpretability, and safety.

\subsection{{\rr{Human-Centred Foundations for AI-Assisted Diagnosis}}}
{\rr{

Human–AI interaction research has long argued that machine learning systems should be designed around the roles that humans play as teachers, evaluators and decision‑makers rather than as passive recipients of predictions.
Amershi et al. noted that effective human supervision requires controllable update mechanisms and interpretable intermediate representations~\cite{Amershi2014Power}. 
Subsequent work synthesized relevant experience into general guidelines for human–AI interaction, emphasizing that AI systems should surface uncertainty and allow users to readily correct or override system outputs~\cite{Amershi2019Guidelines}. 
Human-centered AI frameworks further contend that system design should support both high levels of human control and automation, enabling AI to augment rather than replace professional judgment in safety-critical settings~\cite{Shneiderman2020HCAI}.

Another HCI thread examines how AI systems are embedded in sociotechnical practice. 
Dourish critiqued the expectation that empirical studies must yield explicit ``design implications'', arguing instead for attention to how technologies reshape organizational routines, accountability structures, and professional expertise~\cite{Dourish2006Implications}. 
Greenberg and Buxton observed that when usability evaluation overemphasizes short-term efficiency at the expense of long-term innovation and practice, it may lead to negative effects~\cite{Greenberg2008Usability}. 
Within clinical AI, these perspectives have been extended further: 
Morey et al. argued that evaluation should focus on the joint performance of physicians and AI rather than assessing the model or the user in isolation~\cite{Morey2025Evaluation}, 
while Sokol et al. highlighted AI’s supportive role in diagnostic reasoning~\cite{Sokol2025ClinicalReasoning}. 
Patient-facing studies also show that the guidance style and pacing of chatbot-based family-history collection can influence usability, engagement, and data quality~\cite{Nguyen2024ChatbotFHx}.

Overall, existing research provides an important foundation for the design and evaluation of human-centered diagnostic AI, yet there remains limited systematic exploration of multi-stage diagnostic workflows, particularly end-to-end systems that span history taking, clinical reasoning, and ongoing medical-knowledge updating. 
To this end, we developed and conducted an initial evaluation of a prototype dual-user diagnostic-assistance system that integrates guided conversational history taking \cite{liu2022personal,liu2023low}, LLM–KG–Expert collaborative reasoning, and evidence visualization within a unified workflow aimed at improving the diagnostic experience for both patients and physicians.

}}

\section{Design Requirements and System Overview}

\subsection{Design Requirements}
\rr{
DiagLink aims to optimize the diagnostic process between patients and physicians: on the one hand, to reduce patients’ communication burden and help them understand their own condition; on the other hand, to lower physicians’ cognitive load and time costs so that they can focus more on key diagnostic decisions.
In the context of current medical resource shortages, to understand the specific requirements of patients and physicians, we conducted a literature review at the beginning of the study and held regular discussions with our industry partners, two clinical medical experts each with more than eight years of experience, who were involved throughout the process from prototype iteration to usability validation.
The design requirements (\textbf{DRs}) are detailed below.

\textbf{DR1:Dual-User Centered Design for Asymmetric Collaboration with Clarified Accountability.}
Support collaborative diagnostic workflows between patients and physicians through a unified yet role-adaptive pipeline, while explicitly addressing power, responsibility, and informational asymmetry\cite{oh2022_pgd_asymmetry}, thereby moving beyond the notion of “two separate UIs”.

\begin{itemize}
    \item \textbf{For Physicians (validators and decision-makers).} 
    The interface is designed around diagnostic efficiency and depth, enabling physicians to quickly obtain structured evidence and smoothly complete a physician-led verification process\cite{asgari2024_ehr_cognitive_load}. The initial differential diagnosis list is jointly generated by the LLM and KG but is formally handed over to the physician as soon as they open the case; the physician may revise, supplement, or reject any diagnostic hypothesis, with all changes timestamped and traceable, and final diagnostic responsibility is confirmed and assigned when the physician signs the diagnostic report\cite{bleher2022_diffused_responsibility}.

    \item \textbf{Patient-facing (information providers and collaborators).}
    The interface emphasizes accessibility and guidance, presenting medical information in plain language, assisting symptom description, and encouraging patients’ meaningful collaborative participation\cite{zhang2020_lab_results}. The system does not display any intermediate diagnostic results to patients; all diagnostic information is communicated only by physicians, and only at an appropriate time of their choosing.

    \item \textbf{Across Roles.} 
    Core medical concepts and findings are kept consistent across patient input and physician review to ensure continuity and mitigate misinterpretation~\cite{Dourish2006Implications}.
\end{itemize}

\textbf{DR2: Integrated Human-AI-KG Collaborative Reasoning Framework with Continuous Learning Core Objective.}
Establish a closed-loop reasoning process that synergizes LLMs, KGs, and clinical experts, securing clinical utility and long-term adaptability while embodying the human-centered AI principle~\cite{Shneiderman2020HCAI}.

\begin{itemize}
    \item \textbf{Cyclic Iterative Workflow.}
    Patient input and LLM-guided history-taking populate a case profile; KGs retrieve structured knowledge to generate hypotheses; physicians provide oversight, correct errors, and validate conclusions. This workflow positions experts as “teachers” (Amershi et al.,\cite{Amershi2014Power}) who refine the system's performance over time.
    
    \item \textbf{Attributable Contributions.}
    Different source information (\eg, patient-reported symptoms, LLM-inferred hypotheses, KG-retrieved evidence, physician-validated conclusions) is clearly labeled with its source. This enhances transparency and addresses Dourish's\cite{Dourish2006Implications} concern about “black-box” technologies obscuring accountability.

    \item \textbf{Continuous Knowledge Evolution.} 
    New medical knowledge (\eg, emerging disease associations, updated treatment guidelines) is incorporated into the KG via expert validation—ensuring the system evolves with clinical practice and avoids Greenberg \& Buxton's \cite{Greenberg2008Usability} pitfall of static systems becoming obsolete.
    
\end{itemize}

\textbf{DR3: Unified Presentation of Multi-Source Heterogeneous Evidence.}
Integrate and present evidence from unstructured patient histories, structured KG relationships, and LLM-generated rationales in a cohesive manner—supporting comprehensive decision-making while reducing cognitive load, as advocated by Li et al.,\cite{li2023knowledge}.

\begin{itemize}
    \item \textbf{Visual and Functional Linkage.} 
    Entities referring to the same concept (\eg, "migraine" in patient history and KG) are highlighted with a consistent color scheme, enabling cross-source navigation.

    \item \textbf{Diagnostic switching.}
    Users can switch among different diagnoses and evidence sources, and may display them simultaneously to facilitate comparison and verification.

    \item \textbf{Explicit Evidence Categorization.}
    Evidence types are labeled (\eg, "subjective symptom," "objective guideline," "inferred reasoning") to help physicians quickly assess reliability—critical for junior doctors and time-constrained senior clinicians.

\end{itemize}

\textbf{DR4: Perceptually-Optimized Knowledge Graph Layout for Insight Discovery.}
Design KG visualization to reduce cognitive load and facilitate rapid insight discovery—intuitively representing complex medical relationships and diagnostic confidence—while addressing Greenberg \& Buxton's\cite{Greenberg2008Usability} warning against overwhelming users with excessive information.

\begin{itemize}
    \item \textbf{Clinically Salient Layout.}
    Top-3 candidate diagnoses are positioned as core vertices, common symptoms across diagnoses are aggregated in a central highlighted cluster, and relationship strength (based on path length or evidence score) is encoded via line thickness.

    \item \textbf{Focus-Plus-Context Exploration.}
    Selecting a diagnosis or symptom reorganizes the layout to highlight relevant connections while fading out irrelevant nodes—preserving global context to avoid tunnel vision.

    \item \textbf{Dynamic Interactive Elements.}
    Severity-\zzh{relative likelihood} bars are integrated with the KG view, allowing physicians to adjust thresholds (\eg, "show only high-severity conditions") and tailor visual complexity.
\end{itemize}

}

\subsection{System Overview}

\begin{figure*}[t]
    \centering
    \includegraphics[width=1\linewidth]{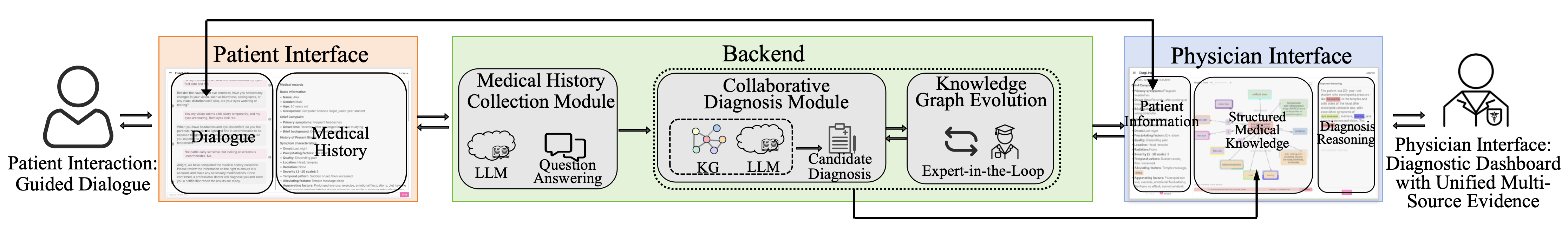} 
    \caption{\name~synergizes LLMs, KGs, and clinical experts to support a dual-user (patients and physicians) diagnostic workflow. It consists of three core modules: (1) a Dialogue-Driven Medical History Collection Module that interactively gathers structured patient data; (2) a Collaborative Diagnosis Module where LLMs and KGs perform parallel reasoning and retrieval-augmented ranking to generate candidate diagnoses, all under a Physician-in-the-Loop mechanism for validation and continuous KG evolution; and 
    (3) a Dual-User Interface Module that provides an \rr{emotionally supportive guided dialogue} interface for patients and an evidence-rich, KG-visualization dashboard for physicians. Arrows indicate the primary flow of data and reasoning, also represent expert-driven feedback and knowledge updates, forming a closed-loop collaborative framework.}
    \label{System Overview}
    \Description{A diagram illustrating the three modules of the system. On the left is the medical history collection module, in the center is the collaborative diagnosis module, and on the right is the user interface with separate patient and physician sides. Arrows indicate the interactions among modules.}
\end{figure*}

\name\ comprises three main modules: a medical history collection module, a collaborative diagnosis module with KG evolution, and a user interface (as illustrated in Fig.~\ref{System Overview}). 
The medical history collection module leverages the powerful natural language processing capabilities of LLMs in combination with template-based constraints, providing patients with a friendly and detailed interaction process that yields complete medical histories. The collaborative diagnosis module operates through the joint efforts of medical experts, LLMs, and KGs, generating the top three most probable candidate diagnoses for each case while iteratively updating knowledge to ensure the system’s continuous refinement and self-improvement. The user interface consists of both patient and physician sides, enabling patients to interact with the medical history collection module, physicians to engage with the collaborative diagnosis module, and also facilitating direct communication between patients and physicians, thereby offering an intuitive and efficient user experience.

\section{System Backend}

This section introduces the backend design of \name\ (DR1-DR3). 
To meet the design requirements outlined earlier, the backend consists of three stages. 
First, patient histories are elicited through guided dialogues. In contrast to prior work that fine-tunes LLMs for direct diagnosis~\cite{tu2025towards}, our approach employs targeted questions to produce semi-structured records. 
Second, in collaborative diagnosis, the system integrates LLMs, KGs, and medical experts : LLMs provide reasoning and generation capabilities, KGs supply structured medical knowledge, and experts dynamically enrich and revise the graph to ensure traceable knowledge evolution and clinical applicability. 
Finally, the system supplements the diagnosis with rationales, treatment recommendations, and supporting evidence from the KG, thereby improving explanability and clinical reliability.

\begin{algorithm}[t]
\caption{Diagnosis-driven Structured History Taking}
\label{alg:history}
\begin{algorithmic}[1]
\REQUIRE mainTemplate, otherTemplate
\ENSURE History, preliminaryDdx
\STATE state $\gets$ \textsc{Main}; messages $\gets \emptyset$; ddx $\gets \emptyset$
\STATE $p \gets$ Greeting()
\WHILE{state $\neq$ \textsc{Done}}
  \STATE \textbf{output: } $p$
  \STATE $u \gets$ GetUserInput(); messages $\gets$ messages $\cup \{u\}$
  \IF{state = \textsc{Main}}
     \STATE ($p$, mainTemplate) $\gets$ UpdateByDialogue(messages, mainTemplate)
     \IF{Complete(mainTemplate)}
        \STATE state $\gets$ \textsc{Other}
     \ENDIF
  \ELSIF{state = \textsc{Other}}
     \STATE ($p$, otherTemplate) $\gets$ UpdateByDialogue(messages, otherTemplate)
     \IF{Complete(otherTemplate)}
        \STATE ddx $\gets$ GeneratePreliminaryDdx(mainTemplate, otherTemplate)
        \STATE state $\gets$ \textsc{Ddx}
     \ENDIF
  \ELSIF{state = \textsc{Ddx}}
     \IF{StoppingCriteria(ddx, messages)}
        \STATE state $\gets$ \textsc{Done}
     \ELSE
        \STATE ($p$, mainTemplate, otherTemplate, ddx) $\gets$ TargetedUpdate(messages, mainTemplate, otherTemplate, ddx)
     \ENDIF
  \ENDIF
\ENDWHILE
\STATE History $\gets$ \{mainTemplate, otherTemplate\}
\STATE preliminaryDdx $\gets$ TopK(ddx, 3)
\RETURN (History, preliminaryDdx)
\end{algorithmic}
\end{algorithm}

\subsection{Dialogue-Driven Medical History Taking}
\label{phase1}

In this stage, \name \ leverages guided dialogues to transform patients’ natural language descriptions into semi-structured medical history records. To this end, we introduce a template-driven mechanism combined with human-centered design principles, thereby ensuring the accuracy and completeness of data collection while simultaneously enhancing patients’ interaction experience.

The medical history template is designed with reference to the CARE guidelines~\cite{riley2017care}, covering key aspects such as patient information, chief complaints, clinical findings, patient perspective, and intervention and outcome (see Supplementary Materials). 
The rationale for employing this template rests on three principles: explicit slot constraints to limit the LLM’s generation scope, thereby mitigating improper inferences and hallucination risks; coverage checking to prevent omission of critical aspects of medical history; and progressive completion to help patients intuitively perceive diagnostic progress.

As shown in Algorithm~\ref{alg:history}, the history-taking process consists of three stages. 
First, the system collects the chief complaint and present illness (\term{Main}), where the LLM generates guided questions based on the dialogue history and template gaps until all relevant information is completed. 
Second, it collects additional histories (\term{Other}), including past, personal, and family histories. 
Finally, in the diagnosis-driven stage (\term{Ddx}), the LLM generates preliminary differential diagnoses, from the collected history and poses targeted questions based on the likelihood and distinctiveness of candidate diagnoses to enrich the record. 
Patient responses are used to instantly update and reorder the diagnosis list. 
The process terminates (\term{Done}) when the candidate diagnoses converge to three results or the maximum number of questions in the \term{Ddx} stage is reached, yielding a complete semi-structured history and a ranked preliminaryDdx.

In each dialogue turn, the LLM determines the most appropriate question by integrating the \textit{messages} with the collected medical history (and the \textit{ddx} during the \term{Ddx} stage). The question is then refined by the LLM to improve its clarity and interaction quality. This process follows the following principles:
it manages cognitive load by limiting each turn to one or two relevant questions, thereby preventing information overload and fatigue~\cite{Sweller2023CLTExpansion}; it employs \rr{emotionally supportive} communication, phrasing questions concisely and naturally to convey respect and warmth, and introducing contextual framing before addressing sensitive topics~\cite{Ayers2023JAMAIM}; and it follows progressive disclosure, moving from general to specific prompts to elicit key information~\cite{bo2024incremental}.

\subsection{Collaborative Diagnosis}
Collaborative diagnosis constitutes the core stage of \name, serving as the bridge between patient histories and diagnostic outcomes. 
In this stage, the joint reasoning of LLMs and KGs broadens diagnostic coverage and sharpens precision, while expert involvement ensures the continuous enrichment and evolution of medical knowledge, thereby enhancing the system's adaptive capacity. 
Collectively, these mechanisms safeguard clinical reliability and secure the system's long-term practical utility.

\subsubsection{Collaborative Diagnosis via LLMs and Evolving KGs}
\label{421}
Previous studies on RAG show that integrating KGs can markedly strengthen the diagnostic capacity of LLMs~\cite{wen2023mindmap, zhao2025medrag}, yet models often become vulnerable to noise when overly dependent on external retrieval ~\cite{baek2023knowledge}. 
Inspired by ~\cite{he2016deep}, we adopt a dual-track strategy: the LLM first produces diagnoses using its internal knowledge, while the KG subsequently generates parallel results through structured reasoning, which are then combined. 
The system subsequently verifies the completeness and currency of relevant KG content, with experts intervening to update and supplement the knowledge base when required (as detailed in Section ~\ref{phase2.2}). 
Finally, adopting a RAG-style paradigm~\cite{lewis2020retrieval}, the system leverages external knowledge from the KG to assist the LLM in ranking candidate diagnoses, from which the top-3 are returned.

\textbf{KG Structured Reasoning.}
Since the preliminary diagnosis from the LLM has been obtained in Section ~\ref{phase1}, we proceed with structured reasoning over the KG. First, the LLM extracts symptom entities from patient histories:
\begin{align}
    \mathcal{E}_{s} = \{s_1, s_2, \dots\} = LLM (Prompt_{recog}, History),
\end{align}
where $\mathcal{E}_{s}$ denotes the recognized symptom set. Each recognized symptom $s_i$ is then linked to the corresponding node in the KG through dense retrieval by encoding it into an embedding and comparing it with all entity nodes $e_j \in \mathcal{E}_g$. The node with the highest similarity is selected if the score exceeds a threshold $\epsilon_s$:
\begin{gather}
     \hat{e}_i = \arg\max_{e_j \in \mathcal{E}_g} \text{sim}\! \ (\text{emb}(s_i), \text{emb}(e_j)), 
     \label{align} \\
     \hat{\mathcal{E}}_s = \left\{\, \hat{e}_i \;|\; 
     \text{sim}\ (\text{emb}(s_i), \text{emb}(\hat{e}_i)) \geq \epsilon_s \,\right\},
\end{gather}
where $\text{emb}(\cdot)$ is the embedding model, $\text{sim}(\cdot,\cdot)$ is the similarity function, $\hat{e}_i$ is the matched entity, and $\hat{\mathcal{E}}_s$ is the set of matched entities. To improve efficiency, we adopt an open-source library FAISS~\cite{douze2024faiss} for large-scale similarity search.

Having obtained the matched symptom entities $\hat{\mathcal{E}}_s$, we then generate candidate diagnostic diseases by exploring their local neighborhoods in the KG. 
For each symptom entity $\hat{e}_j \in \hat{\mathcal{E}}_s$, we retrieve its one-hop neighboring nodes of type disease, forming the candidate set $\mathcal{D}_g$. To refine these candidates, inspired by ~\cite{jia2024medikal}, we incorporate path-based information. For each disease $d_i \in \mathcal{D}_g$, we calculate its shortest path distance to every symptom $\hat{e}_j$, denoted as $dist(d_i, \hat{e}_j)$, defined by the number of edges between them. The relevance score of $d_i$ is then accumulated by summing the reciprocal of path lengths:
\begin{equation}
\text{score}(d_i) = \sum_{\hat{e}_j \in \hat{\mathcal{E}}_s} \frac{1}{dist(d_i, \hat{e}_j)} ,
\end{equation}
where unreachable pairs contribute zero. Finally, all candidate diseases are ranked based on their scores, and the top three results are combined with the preliminary LLM diagnoses to form the combined disease set $\mathcal{D}_{comb}$.

\textbf{RAG-Based Diseases Ranking.}
For each disease $d_i \in \mathcal{D}_{comb}$, we first align it with the corresponding KG node according to Formula~\ref{align}. Diseases originating from the KG are naturally aligned, whereas LLM-generated ones require matching. All $d_i$ can be successfully linked, as missing entities are supplemented by the mechanism described in Section ~\ref{phase2.2}. The aligned set is denoted as $\hat{\mathcal{D}}_{comb}$. 
For each $\hat{d}_i \in \hat{\mathcal{D}}_{comb}$, we then retrieve knowledge $\mathcal{K}$, consisting of three types, from the KG:
(i) its definition; (ii) one-hop symptom neighbors filtered by maximum similarity with $\hat{e}_j \in \hat{\mathcal{E}}_s$; (iii) shortest paths to all $\hat{e}_j \in \hat{\mathcal{E}}_s$, where path information is reused for KG-derived diseases and newly computed for LLM-derived ones. This knowledge, together with the patient’s history, is incorporated into our carefully designed prompt templates (as detailed in Supplementary Materials), enabling the LLM to assign a likelihood score (0–10) to each $\hat{d}_i \in \hat{\mathcal{D}}_{comb}$ and subsequently rank them accordingly.
To ensure concise while reducing physicians’ cognitive load, we retain only the top-3 ranked diseases as the outputs:
\begin{equation}
    \mathcal{D}_{final} = TopK (LLM(Prompt_{rank}, History, \mathcal{K}, \mathcal{D}_{comb}),\; 3).
    \label{eq5}
\end{equation}

\subsubsection{Expert-in-the-Loop}
\label{phase2.2}

{\rr{\name~ treats experts as active agents who iteratively reshape the KG and thereby steer the diagnostic workflow. Each session produces disease predictions (Sec.~\ref{421}) while exposing gaps or obsolete content in the KG, surfaced in the interface (Fig.~\ref{expert}). Experts then select local regions to revise and specify how new knowledge is encoded, forming a closed loop “diagnostic session $\rightarrow$ graph evolution $\rightarrow$ subsequent sessions,” in which direct edits to the KG constrain the system’s future diagnostic behavior.}}

\rr{\textbf{Triggering Expert Engagement.}
\name~ maintains the KG in a disease-centered manner. When constructing the combined disease set $\mathcal{D}_{comb}$ from the LLM and the KG (Sec.~\ref{421}), it inspects the subgraph associated with each disease $d_i \in \mathcal{D}_{comb}$.} An evolution event is triggered if (1) the disease is absent from the KG, (2) its subgraph has never been used in previous cases, or (3) the time elapsed since the last expert-reviewed evolution event exceeds a configurable threshold. \rr{All triggered events are aggregated into a knowledge task worklist for experts.}

\textbf{Disease Subgraph Generation.}
\rr{For each labeled disease $d_i$, \name~updates its subgraph through a semi-automatic “LLM draft + expert finalization” workflow.} Conventional KG completion methods, which rely on fixed schemas or handcrafted rules, struggle to keep pace with rapidly evolving medical knowledge\cite{marchesin_silvello-vldb2024}. \rr{We design a two-stage procedure in which the LLM serves as a drafting assistant and experts act as the final authors of the knowledge representation. Conditioned on the disease name and, when available, its one-hop neighbors in the current KG, the LLM is prompted to generate disease-specific text. This generation is constrained by templates that foreground clinically salient categories, including disease definitions, core and red-flag symptoms, typical clinical course, and first- and second-line treatments. Experts then review the draft in the interface (Fig.~\ref{expert}b) to: (1) correct factual errors, (2) enrich underspecified content (e.g., adding rare but safety-critical manifestations), (3) re-balance the emphasis placed on particular symptoms or therapies, and (4) determine whether the description is sufficiently complete for submission. Through these edits, experts enforce task-level representational constraints, such as clearly separating indications from contraindications and typical symptoms from red-flag features. Once approved, \name~uses targeted prompts to extract structured triples from the curated text and assemble the generative, disease-specific subgraph $\mathcal{G}_s^+$. Since extraction operates on expert-validated text, the resulting edges encode not only factual medical relations but also the priorities and safety considerations embedded in expert editing.}

\textbf{Redundancy and Consistency Checking.}
\rr{To avoid ineffective expansion and contradictions, \name~verifies redundancy and consistency before integrating $\mathcal{G}_s^+$ into the KG. For each generated triple $t = (e_s, r, e_o) \in \mathcal{G}_s^+$, we first discard $t$ if it is a duplicate of an existing triple in the KG. We then calculate its semantic similarity to each existing triple $t' = (e_s', r', e_o')$; if $\text{sim}(t, t') \ge \epsilon_t$, $t$ is treated as a near-duplicate and removed. The remaining triples are merged into the disease-centered subgraph. Experts can then intervene at the graph-level. The interface highlights newly added or modified entities and edges (Fig.~\ref{expert}a) while dimming existing nodes to preserve context, enabling experts to delete or relabel triples or introduce additional links. This graph-level editing enforces higher-order consistency constraints and directly shapes the structure that governs subsequent KG reasoning.}

\begin{figure*}[t]
    \centering
    \includegraphics[width=0.98\linewidth]{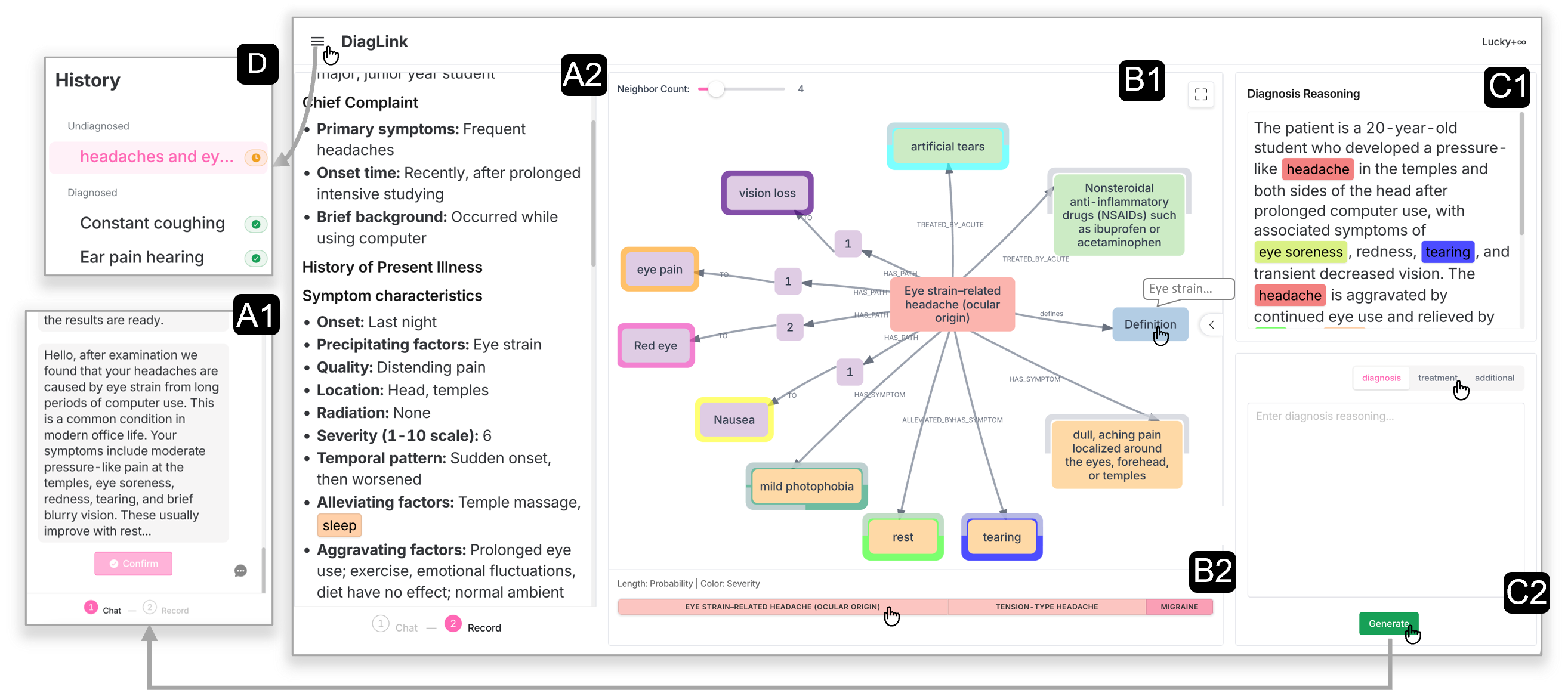} 
    \caption{\textbf{Physician interface}, consisting of three components: patient information (A), medical knowledge (B), and diagnostic reasoning (C). The interface uses a flat layout with highlighted linked entities to integrate heterogeneous data, allowing physicians to quickly understand patient status and diagnostic context. Direct communication with patients is also supported when needed.}
    \Description{A diagram of the physician interface showing three main sections: A) patient information, B) medical knowledge, and C) diagnostic reasoning. The flat layout emphasizes connections between related entities, helping physicians efficiently interpret patient data and reasoning results. Arrows indicate interactions among sections, and optional communication with patients is possible.}
    \label{physician}
\end{figure*}

\rr{\textbf{Closed-Loop Behavioral Feedback.}
Once validated, the disease-specific subgraph is merged into the global KG, where its impact on subsequent diagnoses is mechanistically direct and transparent. Specifically, it: (1) updates the neighborhood of $d_i$ used in KG-based reasoning (Sec.~\ref{421}); (2) supplies the retrieval-augmented ranking module (Eq.~\ref{eq5}) with more complete, up-to-date relations among definitions, symptoms, and medications, improving likelihood estimates and explanations for cases involving $d_i$; and (3) is exposed in the clinician view (Fig.~\ref{physician}), together with metadata such as the reviewing expert and timestamp. As more diseases traverse this loop, fewer cases trigger evolution, and experts shift from high-frequency annotators to low-frequency supervisors and strategy designers, while the system’s diagnostic behavior co-evolves with the KG. Thus, expert interventions systematically reshape the KG and, in turn, how \name\ generates, ranks, and explains diagnoses, establishing an expert-in-the-loop feedback cycle.}

\subsection{Evidence-Based Enhancement}

Beyond enhancing the trustworthiness of diagnostic outcomes through collaborative mechanisms, this section further focuses on consolidating and generating evidence to help physicians better understand and \rr{assess} candidate diagnostic conclusions, while simultaneously alleviating their cognitive burden during decision-making.
For each candidate disease $d_i \in \mathcal{D}_{final}$, the system first retrieves supporting information from the KG, including the disease definition, one-hop symptom neighbors, the shortest paths connecting patient symptoms, one-hop drug neighbors, and their extended neighbors. 
Nodes of different categories are further ranked by importance, enabling physicians to selectively expand an appropriate number of nodes as needed.
Specifically, for all triples except disease definitions, the system prompt the LLM to rank them according to their relevance to both the target disease and the patient history, while strictly preserving the original content. Building on this, the system integrates patient history with the retrieved evidence to refine the diagnosis, prompting the LLM to generate logically coherent and concise reasoning as well as clinically appropriate treatment recommendations.

\section{Visualization and Interaction}

\rr{This section presents the visualization and interaction design of \name\ (DR1-DR4). 
for patients and physicians. Since the expert interface and its interactions have been described in Sec.~\ref{phase2.2}, our focus will be here on the patient- and physician-facing interfaces and interactions.}

\begin{figure}[t]
    \centering
    \includegraphics[width=0.98\linewidth]{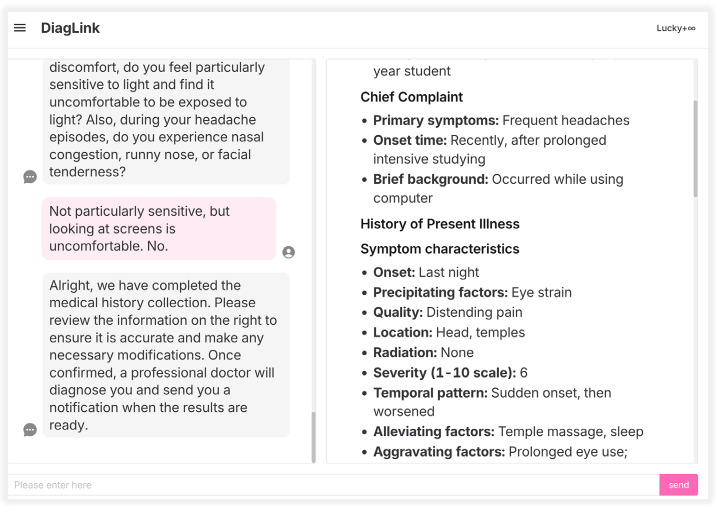}
    \caption{\textbf{Patient Interface.} \name\ provides a simple and intuitive conversational interface for collecting medical history while dynamically displaying the collected information on the right.}
    \label{patient}
    \Description{This interface enables clinicians to collect patient data through natural language dialogue while simultaneously displaying and updating the patient's medical history in a dedicated panel for context-aware decision-making.}
\end{figure}

\begin{figure}[t]
    \centering
    \includegraphics[width=0.98\linewidth]{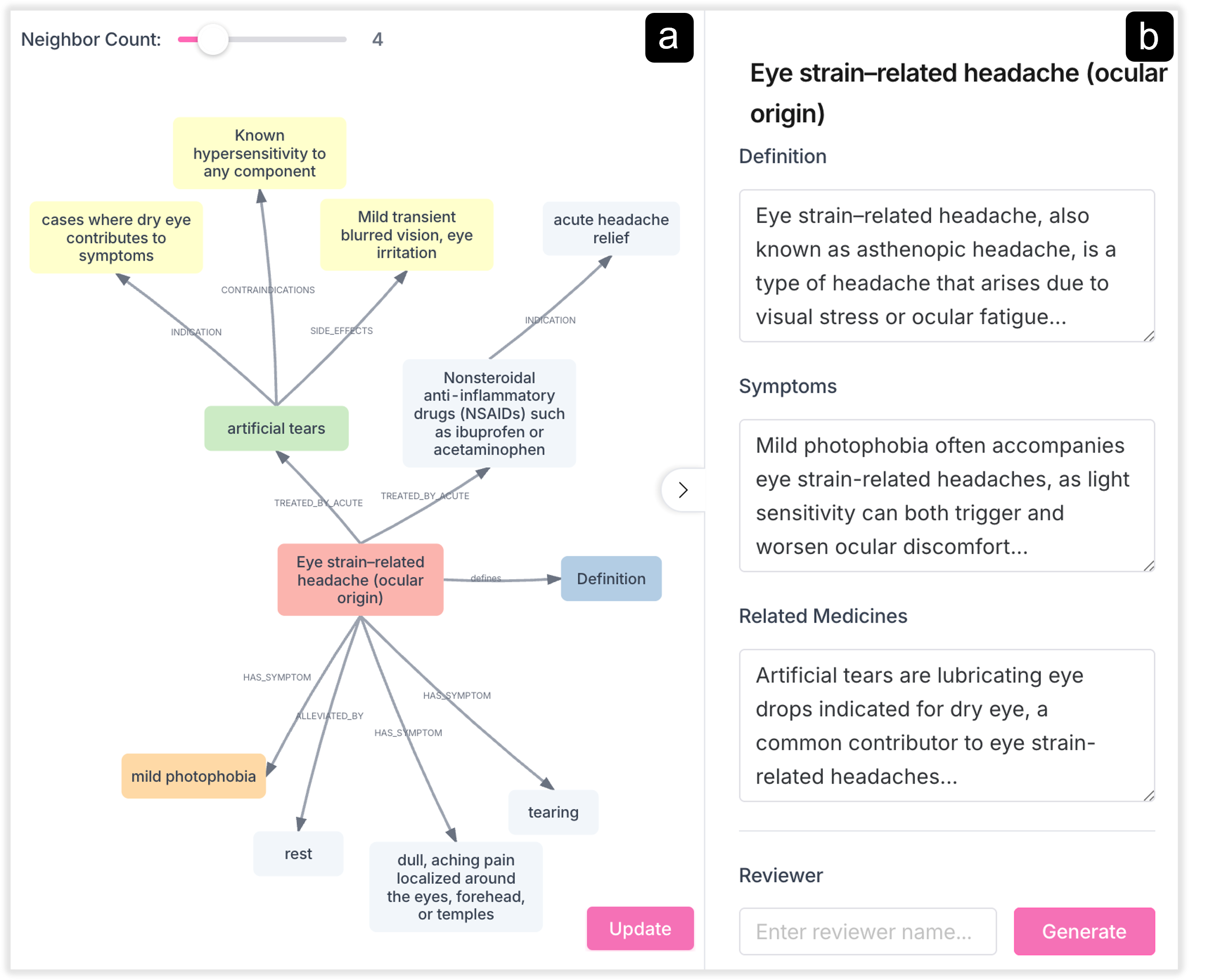} 
    \caption{\textbf{Expert Interface.} Expert can iteratively update, validate the \name's knowledge. Newly added or modified components are highlighted to emphasize recent changes, while previously entities remain visible in a subdued form. This design supports focused editing while preserving awareness of the overall structure.}
    \label{expert}
    \Description{The interface provides an integrated environment where experts can progressively update, refine, and manage system knowledge. Recent modifications are highlighted for clarity, while stable elements are shown in a faded state to maintain context.}
\end{figure}

\rr{\subsection{Architecture}}

The system consists of three main interfaces, the patient interface (Fig.~\ref{patient}), the physician interface (Fig.~\ref{physician}), and the internal supplementary expert interface (Fig.~\ref{expert}).
\rr{The patient interface provides a dialogue panel and a dynamic, semi-structured history panel. Within this interface, entity mentions (e.g., symptoms, diseases, and medications) are recognized and normalized to KG concepts.
The physician interface contains a three-part dashboard: (1) patient information; (2) diagnostic KG; (3) reasoning and recommendation. These three parts are linked; selecting an entity highlights its counterparts across panels.
The expert interface offers a task worklist, a subgraph editor, and a diff view. Newly added or modified nodes/edges are highlighted and each commit to the KG is attached with provenance and a timestamp.}

\subsection{Patient interaction}
Based on consultations with domain clinical experts, we identified that patients often often struggle to accurately describe symptoms. 
\rr{To address this, \name~uses an LLM–driven conversational interface to collect medical history through guided, supportive dialogue. Patients describe their concerns in natural language, answer follow-up questions, and verify and revise information as prompted. 
A history Panel continuously aggregates and updates key details into a clear semi-structured record for clinicians. 
For the interaction process, after login, patients first receive a brief greeting and are then invited to describe their current problems; based on the accumulated information, the system adaptively generates follow-up questions and, for sensitive topics, prefaces these questions with a brief explanation of their purpose to reduce discomfort.
Once core history has been obtained, the system constructs preliminary diagnostic hypotheses on the backend and derives discriminative questions to elicit targeted additional information. After all questions are completed, patients review and amend the summarized history before submission. The system then confirms completion and informs patients that physicians will perform the diagnosis. Any updates will be delivered via notifications, eliminating the need for patients to wait continuously. Throughout, no intermediate diagnostic hypotheses or inferences are shown to patients; all diagnostic information is communicated only by clinicians.}

\subsection{Physician Interaction}
{\rr{From interviews with medical experts, the physician-facing interface is designed to shorten decision time without compromising safety, reduce the cognitive load of information retrieval and integration, and support reliable decisions through an explainable reasoning process grounded in traceable knowledge. Clinicians emphasized that a bare diagnostic output is inadequate; the system must expose its evidence structure and reasoning, and make it easy to reach a complete diagnosis.
To this end, \name~ reduces the interface to three core components: patient information (Fig.~\ref{physician}A), medical knowledge (Fig.~\ref{physician}B) and diagnostic reasoning with recommended treatment (Fig.~\ref{physician}C). A flat three-column layout highlights identical entities across panels, bringing together heterogeneous information from dialogue logs, the KG, and the reasoning engine in a single view. 
\begin{itemize}
    \item (A) patient information: clinicians can switch between the raw dialogue and a semi-structured history, and, when needed, continue the conversation with the patient to obtain additional details.
    \item (B) medical knowledge: it built on a verified KG, automatically selects objective evidence most relevant to the current candidate diagnoses and orders it by clinical importance or strength of evidence. Hover actions reveal metadata such as review time and reviewer identity, enabling rapid assessment of reliability and timeliness. To support comparison, \name~ displays only the top three candidate diagnoses. In (B2), bar length encodes the system-estimated \zzh{relative likelihood} and colour intensity encodes the potential severity of outcomes if the disease is missed; hovering shows exact values. When the active diagnose changes (B2), (C) update synchronously.
    \item (C) diagnostic reasoning with recommended treatment: it presents structured reasoning steps and treatment suggestions automatically derived from the evidence in (B), making the reasoning process visible and auditable. Clinicians can directly edit, add or remove steps and management items, explicitly injecting their experience and contextual judgement to correct biases or add necessary detail. 
\end{itemize}
When they confirm that a candidate diagnosis is final, they select its entry in (B2) and complete or adjust a small number of key fields (e.g., diagnostic conclusion, treatment plan, follow-up schedule and precautions). \name~ then rewrites the clinician-approved diagnosis and plan into a patient-facing explanation, which is sent to the patient interface (A1) after final review. 
By tightly coupling interpretable reasoning, traceable knowledge and clinician judgement, \name~ streamlines workflow and reduces cognitive load while keeping diagnostic safety and responsibility boundaries clear and controllable.
}}

\begin{figure}[t]
    \centering
    \includegraphics[width=0.98\linewidth]{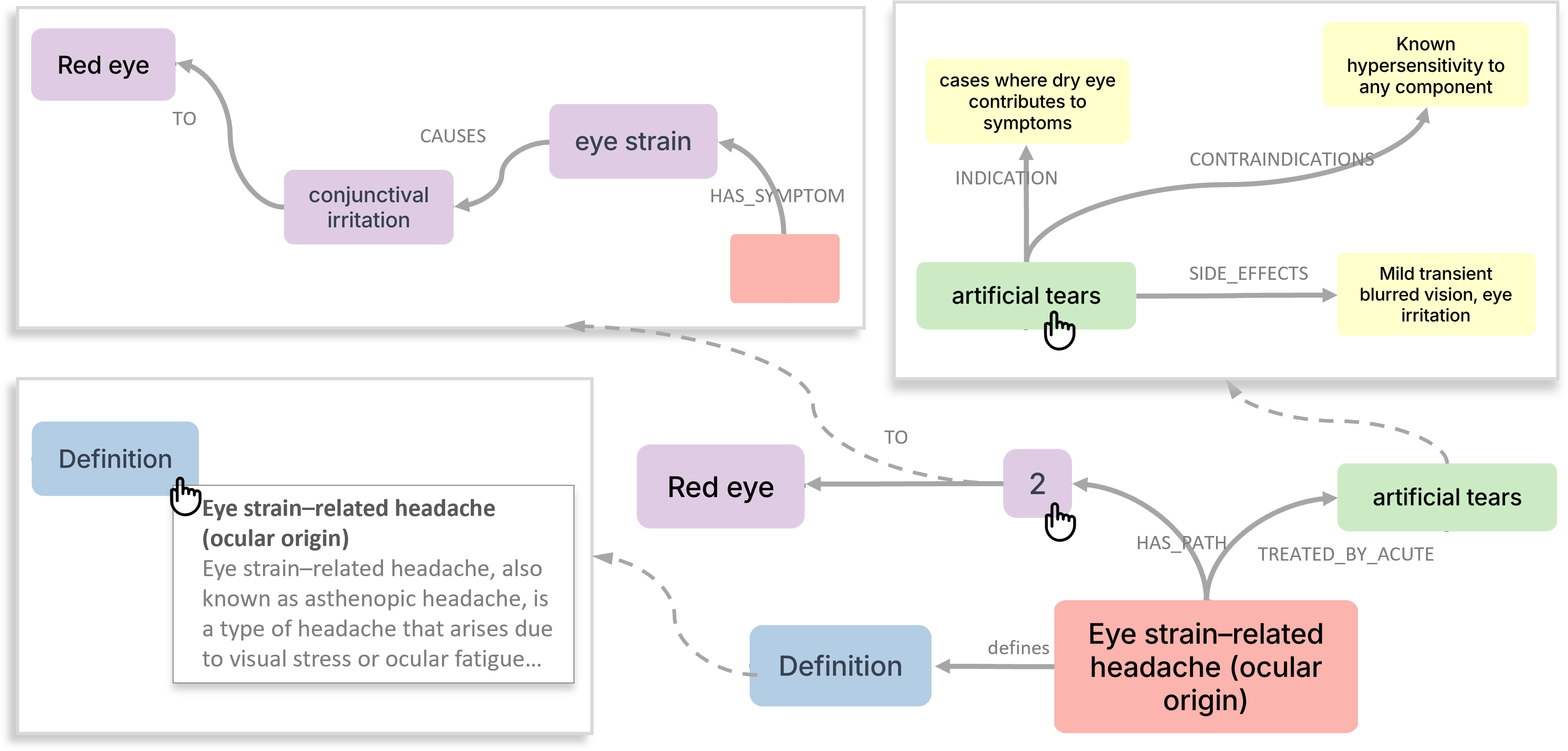} 
    \caption{\textbf{Entity Exploration.} Users can explore knowledge about entities in three forms: definition nodes, drug nodes, and shortest-path nodes connecting diseases and symptoms. This design balances information richness with cognitive load, allowing users to access relevant details without being overwhelmed.}
    \label{detail}
    \Description{The interface supports exploration of entities in three formats: definition nodes, drug nodes, and shortest-path nodes linking diseases and symptoms. It is designed to present information efficiently while minimizing cognitive overload.}
\end{figure}

\subsection{Diagnostic Knowledge Graph View}
Diagnostic Knowledge Graph View (Fig.~\ref{physician}) aims to visualize the relational structure of the medical KG, providing physicians with a global understanding of the diagnostic context. This view supports both diagnostic decision making and expert supplementation. The Diagnostic Knowledge Graph View consists of three parts: a knowledge, a navigating bar, and a knowledge editing box.

To better assist doctors in organizing the relationships between knowledge, we introduce a layout~\cite{feng2020topology} for the visualization of the KG. The result of the top-k diagnosis (in this study we default to k = 3) is arranged as vertices of a polygon (\ie, a triangle when k=3). The KG categorizes entities into five types: candidate diagnoses, one-hop symptoms linked to diagnoses, patient-history-derived symptoms with shortest paths to diagnoses, definitions, and drugs.
One-hop symptoms which common to all candidate diagnoses are aggregated in a central of global view, highlighted in yellow. 
Patient-history-derived symptoms are highlighted in purple. To maintain visual clarity, the number of hops—calculated as the shortest path between disease and symptom entities—is initially summarized numerically. Physicians can click on any symptom node to expand and inspect the actual paths within the graph (Fig.~\ref{detail}).
Definitions are presented in a similar manner in blue, with their detailed content displayed in a tooltip-like bubble only when physicians click on them.
In addition, related drugs are depicted in green, and a neighboring node is displayed when it is clicked.

Below the Diagnostic Knowledge Graph View, a navigation bar (Fig.~\ref{physician} B2) is incorporated to facilitate interactive diagnosis exploration. Each bar within this component corresponds to a candidate diagnosis generated by the system. Physicians can directly click on any bar to select the associated diagnosis result, triggering an immediate update of the KG visualization to focus on the chosen diagnosis result and its related medical entities.
When a physician selects a particular diagnosis result (\eg, \textit{"eye strain–related headache (ocular origin)"} in Fig.~\ref{physician}), the view dynamically focuses on the chosen disease. All unrelated nodes including other candidate diagnoses and irrelevant symptoms—are faded to gray. In this focused mode, the visualization recomputes the layout to enhance interpretability: the selected diagnosis is placed at the center, with associated entities (symptoms, drugs,\textit{ etc}.) grouped categorically and arranged radially around it. This reorganization minimizes clutter while emphasizing connections relevant to the chosen diagnosis.

{\rr{\subsection{Implementation}

\textsc{DiagLink} adopts a web-based client--server architecture using Nuxt.js~\cite{nuxt}/Vue~3~\cite{vue3} with Nuxt-Auth.js~\cite{nuxtauth} on the front end and Flask~\cite{flask} on the back end. Interactive KG exploration is supported by Graphology~\cite{graphology} for graph manipulation and Sigma.js~\cite{sigmajs} for rendering, enabling smooth navigation of dense disease-centered neighborhoods. The KG is stored in Neo4j~\cite{neo4j}. For natural language understanding and clinical reasoning, \textsc{DiagLink} integrates GPT-4.1~\cite{gpt41}, which at the time of development belonged to the SOTA class of general-purpose LLMs available in practice, encapsulated behind a modular interface so that alternative models can be plugged in~\cite{sun2020dfseer} with minimal changes. PrimeKG~\cite{Chandak2023PrimeKG} serves as the main biomedical resource, a multimodal KG built from curated databases and ontologies with roughly 130k nodes and 4M edges, including about 17k diseases and 8k drugs. The framework itself is agnostic to the specific graph and can also operate on alternative or proprietary KGs. 
The source code is publicly available as open source at \url{https://github.com/meguriri/DiaLink}.

}}

\rr{\section{Controlled User Study}}
\label{user}

\rr{To evaluate the usability, effectiveness, and safety of \name~ in enhancing diagnostic experiences for both patients and physicians, we designed and conducted a within-subject controlled study. The study involved 12 clinical scenario packs, 12 simulated patients, and 12 practicing physicians. Each simulated patient was assigned a fixed scenario and, under each of the three system conditions (i.e., \name, Baseline A, and Baseline B), completed one consultation with a different physician.}

\rr{\subsection{Study Setup}}
\subsubsection{Participants and Data}

\rr{We recruited 12 simulated patients (SP1–SP12; 6 male, 6 female; mean age 24.2) and 12 physicians (P1–P12; 5 male, 7 female; mean practice experience 5.3 years) through social media and word of mouth. Physicians reported an average score of 3.8 (5-point Likert scale) for their familiarity with LLM-based chatbots. All simulated patients possessed basic health literacy and prior care-seeking experience, with a mean self-efficacy score of 4.2 (5-point Likert scale) for clinical communication.} \rr{Scenario packs were sourced from the samples provided on the Membership of the Royal Colleges of Physicians UK website\footnote{https://www.thefederation.uk/resources/}. 
Each scenario includes a ground-truth diagnosis and acceptable differential diagnoses.}

\subsubsection{Baselines}

\rr{Two baseline systems were implemented. (1) {Baseline A} provides patients with a text-based interface for conversing with the physician. On the physician side, it offers a text chat interface for patient communication, a KG viewer that retrieves subgraphs for a single queried entity, and an LLM-powered chat interface for diagnostic support. The underlying KG and LLM are identical to those used in \name, and the overall visual design of the interface is matched to \name~ to minimize confounding from aesthetic differences.
(2) {Baseline B} is identical to the full \name~ system except for the timing of presenting system-generated diagnostic predictions. Predictions are shown only after explicit physician confirmation, enabling the examination of potential anchoring effects of system-provided diagnoses on clinical decision-making.}

\subsubsection{Tasks and Experimental Design}
\rr{All tasks were performed in a desktop environment, with each device connected via SSH to a centralized server. A within-subject balanced design was adopted. Each simulated patient used their assigned scenario package and completed one consultation under each of the three system conditions (\name, Baseline A, Baseline B). Each physician worked through 3 distinct scenario packs, distributed across the three conditions. Every physician–patient pairing occurred only once, ensuring that each simulated patient interacted with three different physicians across conditions. System-condition order was counterbalanced to mitigate order effects.}

\subsubsection{Procedure}

\rr{Before the formal experiment, all participants received unified training using an additional scenario that was not included in the analysis. For simulated patients, we used this scenario to explain the requirements of role-playing, including the scope of information disclosure, consistency of performance, and response principles; demonstrations and brief practice were provided to help them master simulated acting. Each simulated patient then individually read and familiarized themselves with their assigned formal scenario. The researcher confirmed readiness before ending this stage, typically within 20 minutes. Physicians used the same additional scenario to become familiar with the interfaces and workflows of the three systems.}
\rr{In the formal study, to balance time burden and facilitate scheduling, the experiment was organized into four rounds. Each round involved three physicians and three simulated patients, and each patient/physician participated in only one round. Within each round, a 3×3 Latin square design was used to counterbalance the assignment of three system conditions across the three physicians and three cases. Three researchers were responsible for coordination and documentation. Physicians were instructed to complete each diagnostic task within 20 minutes.}
\rr{After each consultation, both parties completed corresponding questionnaires.
Simulated patients, under the \name~ and Baseline A conditions, completed questionnaires (Fig. \ref{patient_satisfaction}) using a think-aloud protocol to assess the impact of the system on patient-centered communication experience and revisit intention.
Physicians submitted a ranked list of differential diagnoses for every system condition. Under \name~ and Baseline A, they additionally provided test orders, treatment and management plans, and follow-up recommendations (details in Appendix), and evaluated subjective workload using the NASA Task Load Index (NASA-TLX) questionnaire\cite{Hart1988NASA_TLX}. Under \name, they also completed the System Usability Scale (SUS) questionnaire\cite{Brooke1996SUS} and rated their trust in system-generated diagnostic predictions. The diagnosis time was recorded.}
\rr{For diagnostic and management records collected under \name~ and Baseline A, we invited two medical experts (E1, E2), who independently performed blinded evaluations in randomized order.
Experts had access to the full scenario packs including ground-truth diagnosis and acceptable differential diagnoses. 
Assessment criteria included the
accuracy and appropriateness of differential diagnoses and the appropriateness of investigation, treatment, management plan, and follow-up 
(Table~\ref{expert_q} in Appendix). Discrepancies were resolved through discussion.}

{\rr{

\subsection{Quantitative Results}

\subsubsection{Patient-Side Diagnostic Experience}

To assess patient-centered diagnostic experience, we administered a 5-point Likert-scale questionnaire (Q1–Q10) covering four dimensions: Relationship \& Respect (Q1–Q4), Participation \& Understanding (Q5–Q6), Emotional Support \& Welfare (Q7–Q9), and Revisit Intention (Q10). As shown in Fig. ~\ref{patient_satisfaction} and Table ~\ref{tab:patient-sig}, patient ratings for \name~ are significantly higher than for the baseline system.
In the Relationship \& Respect dimension, patients report higher ratings for ``good rapport and connection with the doctor'' (Q1; Diff = 0.83, p = .0105), with the largest improvement observed for ``relaxed and at ease'' (Q4; Diff = 2.25, p < .001).
In the Participation \& Understanding dimension, ratings are significantly higher than the baseline for both ``encouraged to participate'' (Q5; Diff = 1.00, p = .0261) and ``understand the condition and related information'' (Q6; Diff = 0.75, p = .0001). The semi-structured medical history increases transparency and  gives patients a stronger sense of control during diagnosis.
In the Emotional Support \& Welfare dimension, \name~ increases patients’ perceived care and emotional support. The improvement is particularly notable for “comfort, emotional well-being, and dignity being valued and protected” (Q9; Diff = 1.08, p = .0016). The language optimization support helps patients understand better and feel more relaxed and respected.
In the Revisit Intention dimension, \name~ substantially increases patients’ willingness to use this diagnostic approach again in the future (Q10; Diff = 1.42, p = .0001). The overall improvements translate into greater acceptance of this diagnostic process.

\begin{figure*}[t]
    \centering
    \rrfig[width=0.888\linewidth]{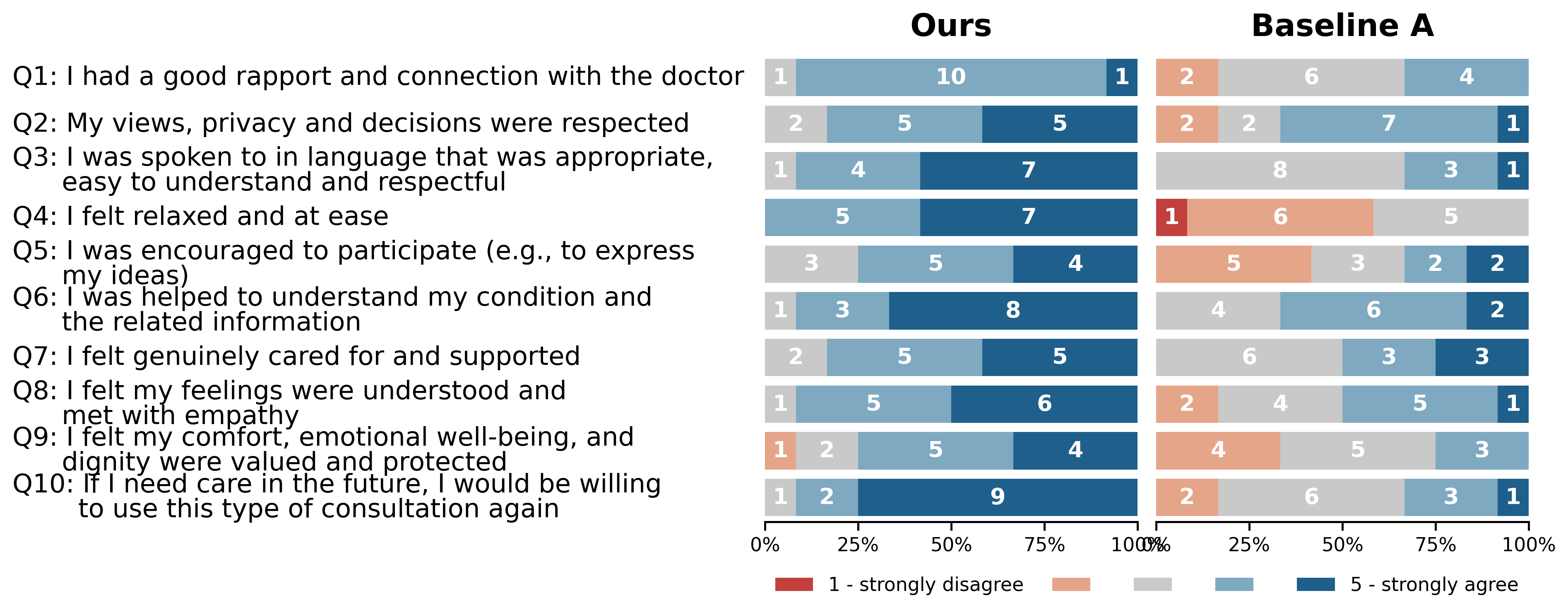}
    \caption{\rr{\textbf{Simulated patient satisfaction ratings for \name~ versus Baseline A on a 5-point Likert scale.} Each row shows the distribution of responses (1 = strongly disagree to 5 = strongly agree) for ten survey items, grouped into four dimensions: Relationship \& Respect (Q1–Q4), Participation \& Understanding (Q5–Q6), Emotional Support \& Welfare (Q7–Q9), and Revisit Intention (Q10).}}
    \label{patient_satisfaction}
    \Description{Stacked bar chart comparing simulated patient satisfaction ratings between \name~ and Baseline A for ten questions on a 5-point Likert scale. For most questions, \name~ has a higher proportion of ``agree'' and ``strongly agree'' responses than Baseline A across all four dimensions.}
\end{figure*}

\subsubsection{Physician-Side Diagnostic Experience}
We assessed physicians’ subjective cognitive workload using the NASA-TLX for both systems and evaluated the usability of \name~ with the SUS. As shown in Fig. ~\ref{fig:nasa_tlx}, \name~ significantly reduced workload across all six NASA-TLX dimensions. The overall weighted workload decreased from 12.46 with the baseline system to 6.75 (p < .001). The largest reductions were observed in Temporal Demand (D3) and Effort (D5), with decreases of 11.5 and 9.8 points, respectively (both p < .001). 
Mental Demand (D1) and Frustration (D6) were slightly higher under \name~ compared with the other dimensions, but still markedly lower than those of the baseline system. 
Improvement in Performance (D4) was more modest (Diff = 1.4, p < .05) but remained statistically significant. In terms of weight distribution, \name~ reduced the contribution of Temporal Demand from 21\% to 10\%, effectively alleviating time pressure and freeing cognitive resources for core tasks. 
Overall, the benefits of \name~ mainly stem from reduced time urgency and operational effort, while also contributing positively to perceived performance. 
Based on the SUS, \name~ achieved a score of 77.5, indicating an overall acceptable level of usability and outperforming approximately 80\% of comparable application systems~\cite{Bangor2008SUS, SauroLewis2016QuantifyingUX}. 
The mean ratings for the positively worded (odd-numbered) items exceeded 4, whereas those for the negatively worded (even-numbered) items remained below 2. This response pattern reflects a consistently favorable user perception of the system’s ease of use.

\begin{table}[t]
\centering
\rrtab{\begin{tabular}{lccccc}
\toprule
\textbf{Dim.} & \textbf{M.(\textit{Baseline})} & \textbf{M.(\textit{Ours})} & \textbf{Diff.} & \textbf{t} & \textbf{p} \\
\midrule
Q1  & 3.17 & 4.00 & 0.83 & 3.08 & .01048 \\
Q2  & 3.58 & 4.25 & 0.67 & 2.97 & .01283 \\
Q3  & 3.42 & 4.50 & 1.08 & 4.17 & .00157 \\
Q4  & 2.33 & 4.58 & 2.25 & 10.34 & .00000 \\
Q5  & 3.08 & 4.08 & 1.00 & 2.57 & .02609 \\
Q6  & 3.83 & 4.58 & 0.75 & 5.74 & .00013 \\
Q7  & 3.75 & 4.25 & 0.50 & 2.57 & .02609 \\
Q8  & 3.42 & 4.42 & 1.00 & 3.32 & .00687 \\
Q9  & 2.92 & 4.00 & 1.08 & 4.17 & .00157 \\
Q10 & 3.25 & 4.67 & 1.42 & 6.19 & .00007 \\
\bottomrule
\end{tabular}}
\caption{\rr{\textbf{Statistical comparison of simulated patient satisfaction scores between Baseline A and \name.} 
The table reports mean scores (M.) on a 5-point Likert scale for each of the ten survey items in Fig.~\ref{patient_satisfaction}, comparing Baseline A and \name. Diff.\ denotes the mean difference (Ours minus Baseline), while \textit{t} and \textit{p} correspond to the paired t-test statistics.}}
\label{tab:patient-sig}
\Description{\textbf{Table summarizing the statistical advantage of \name~over Baseline A on simulated patient satisfaction scores.} Each row corresponds to one questionnaire item (Q1–Q10) shown in Fig.~\ref{patient_satisfaction}. For every item, the mean rating under \name~is higher than under Baseline A, yielding positive differences between 0.50 and 2.25 points on a 5-point Likert scale. The associated paired t-tests indicate that all improvements are statistically significant (p < .05), with the largest effect observed for Q4 (rapport and comfort during the consultation) and substantial gains also present for items related to participation, understanding, empathy, and willingness to use the system again.}
\end{table}

\subsubsection{Diagnostic Performance from the Expert Perspective}

To assess diagnostic performance from the expert perspective, we administered a 5-point Likert-scale questionnaire (Q1–Q7). Among these items, Q1 and Q6 used the original five-level scale, while the remaining items were originally Yes/No questions and were mapped to a five-level scale during analysis. Details on the questionnaire source and the mapping rules are provided in Appendix. Fig.~\ref{expert_r} and Table~\ref{expert-sig} present a comparison between \name~ and Baseline A.
\name~ performs better overall from the expert viewpoint. Except for Inappropriate treatment avoided (Q5), \name~ achieved higher mean scores on all items, with Q1, Q2, Q4, and Q6 reaching varying levels of statistical significance, respectively (p<0.01, p<0.05, p<0.01, p<0.001). \name~ supports the physician decision-making process more effectively, including more structured evidence and more consistent and reasonable recommendations. We further analyzed diagnostic accuracy and completion time of physicians across 12 scenario cases. \name~ increased the number of correct Top-1 and Top-3 cases from 7 and 9 (baseline) to 9 and 11, respectively, and reduced the average diagnostic time from 18.6 minutes to 7.8 minutes. 
Diagnostic correctness was determined based on expert annotation: a candidate diagnosis was considered correct if it matched the ground truth exactly or was judged to be highly similar or closely related (including acceptable differential diagnoses).
Overall, \name~ substantially reduced diagnostic time, improving physician efficiency while also providing a moderate improvement in diagnostic effectiveness.

\begin{figure}[t]
    \centering
    \rrfig[width=\linewidth]{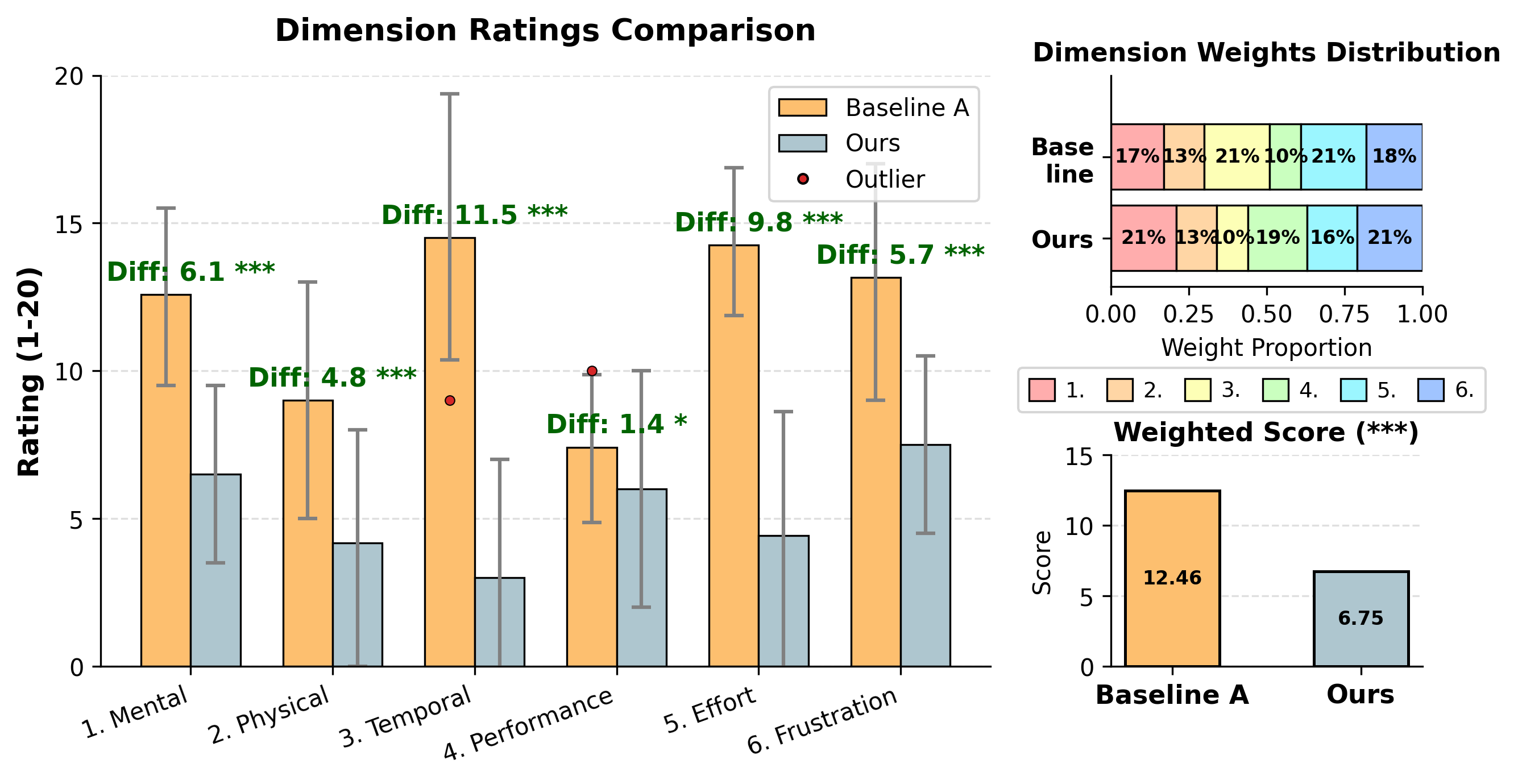} 
    \caption{\rr{\textbf{NASA-TLX workload comparison between Baseline A and \name.} 
    The left panel shows physicians' mean NASA-TLX ratings (1--20) on six workload dimensions for Baseline A (orange) and \name~(blue). Gray error bars indicate the range of ratings after removing outliers using the 1.5$\times$IQR rule, and red dots mark the excluded outliers. The top-right panel depicts the relative weight assigned to each dimension, and the bottom-right panel summarizes the resulting weighted NASA-TLX scores.}}
    \label{fig:nasa_tlx}
    \Description{Figure comparing physicians' perceived workload between Baseline A and \name~using NASA-TLX. The main bar chart contrasts mean ratings and non-outlier ranges across six dimensions, and the right-side panels show the corresponding dimension weight distributions and overall weighted workload scores, with \name~yielding a smaller weighted NASA-TLX score than Baseline A.}
\end{figure}

\begin{figure*}[!t]
    \centering
    \includegraphics[width=0.798\textwidth]{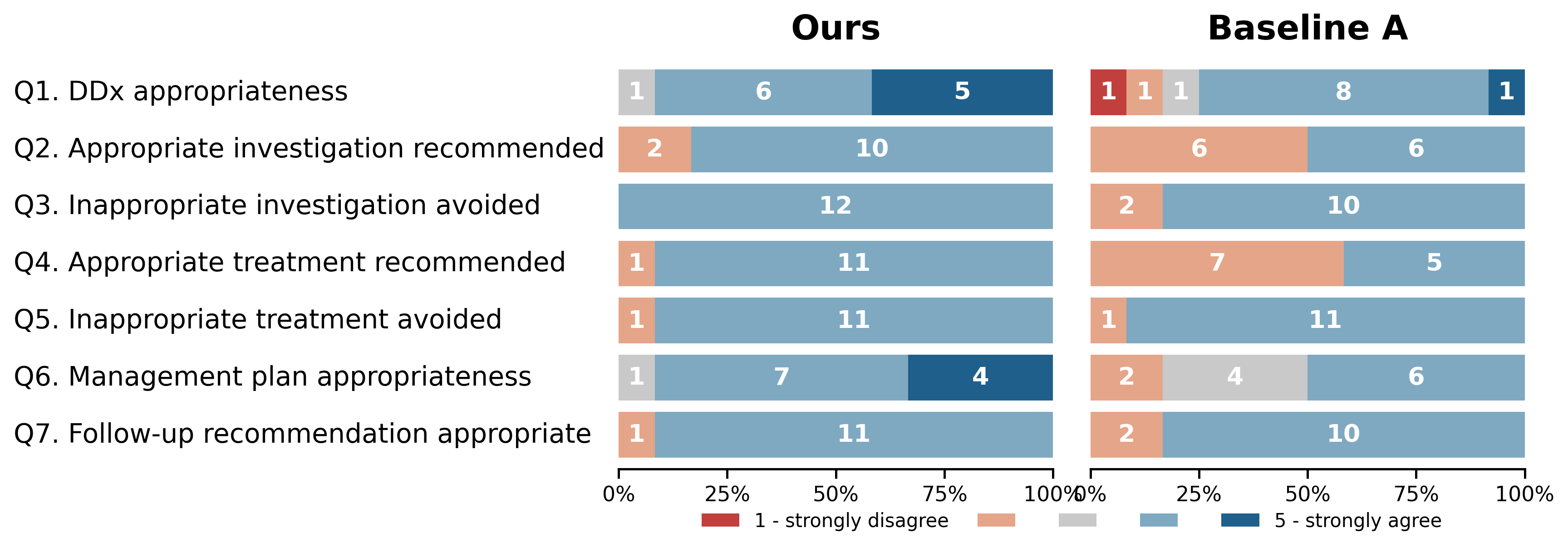} 
    \caption{\rr{\textbf{Expert ratings for \name~ versus Baseline A.} Stacked 5-point Likert bars for seven clinical questions (Q1–Q7) on diagnostic, investigation, treatment, management, and follow-up quality. The full questionnaire items are provided in Appendix Table~\ref{expert_q}.}}
    \label{expert_r}
    \Description{A horizontal stacked bar chart compares expert ratings for two methods, labelled “Ours” and “Baseline A”, across seven clinical questions (Q1–Q7). Each row contains two bars, one for each method, segmented into five coloured sections representing Likert-scale scores from 1 (strongly disagree) to 5 (strongly agree). The bars for \name~ are shifted toward the higher-score segments (4 and 5), while Baseline A shows relatively more lower and mid-range scores, indicating consistently more favourable expert assessments for \name~ across all questions.}
\end{figure*}

\begin{table}[t]
\centering
\rrtab{\begin{tabular}{lccccc}
\toprule
\textbf{Dim.} & \textbf{M.(\textit{Baseline})} & \textbf{M.(\textit{Ours})} & \textbf{Diff.} & \textbf{t} & \textbf{p} \\
\midrule
Q1 & 3.58 & 4.33 & 0.75 & 3.45 & .00546 \\
Q2 & 3.00 & 3.67 & 0.67 & 2.35 & .03881 \\
Q3 & 3.67 & 4.00 & 0.33 & 1.48 & .16609 \\
Q4 & 2.83 & 3.83 & 1.00 & 3.32 & .00687 \\
Q5 & 3.83 & 3.83 & 0.00 & ---  & ---    \\
Q6 & 3.33 & 4.25 & 0.92 & 6.17 & .00007 \\
Q7 & 3.67 & 3.83 & 0.17 & 1.00 & .33880 \\
\bottomrule
\end{tabular}}
\caption{\rr{\textbf{Statistical comparison of expert ratings between Baseline A and \name.} 
The table reports mean scores (M.) on a 5-point Likert scale for each of the seven survey items in Fig.~\ref{expert_r}, comparing Baseline A and \name. Diff.\ denotes the mean difference (Ours minus Baseline), while \textit{t} and \textit{p} correspond to the paired t-test statistics.}}
\label{expert-sig}
\Description{\textbf{Table summarizing the statistical advantage of \name~over Baseline A on simulated patient satisfaction scores.} Each row corresponds to one questionnaire item (Q1–Q10) shown in Fig.~\ref{expert_r}. For every item, the mean rating under \name~is higher than under Baseline A, yielding positive differences between 0.50 and 2.25 points on a 5-point Likert scale. The associated paired t-tests indicate that all improvements are statistically significant (p < .05), with the largest effect observed for Q4 (rapport and comfort during the consultation) and substantial gains also present for items related to participation, understanding, empathy, and willingness to use the system again.}
\end{table}

\subsubsection{Safety}
This section provides a quantitative assessment of the safety of \name~.

\textbf{Anchoring Effect.}
The anchoring effect refers to physicians’ tendency to over-rely on initial information, potentially influencing subsequent judgments. To evaluate its presence in \name~, we conducted a paired comparison between \name~ and Baseline B. Baseline B achieved 10 and 11 correct cases in the Top-1 and Top-3 accuracy scenarios, respectively, with an average diagnostic time of 9.8 minutes. Its Top-3 accuracy was the same as that of \name~, while its Top-1 accuracy was slightly higher, which we speculate may be related to its longer analysis time. A paired binary statistical analysis using the exact two-sided McNemar test indicated no significant difference between the two systems (p = 1.00). Therefore, we conclude that \name~ does not exhibit a noticeable anchoring effect. By jointly presenting patient history, preliminary diagnoses and their underlying reasoning, as well as knowledge from the KG, \name~ supports physicians in forming their judgments. Its emphasis on transparent diagnostic reasoning and evidence presentation—rather than solely displaying final conclusions—helps reduce the risk of the anchoring effect.

\textbf{Diagnostic \zzh{Relative Likelihood} and Physician Trust.} 
To assess whether \name~’s diagnostic \zzh{relative likelihood} scores can differentiate between correct and incorrect diagnoses—and whether physicians’ trust aligns with diagnostic quality—we divided the 12 scenario packs (36 diagnostic samples) into "correct" (17 samples) and "incorrect" (19 samples) groups based on expert annotations (Fig. \ref{box}). Each sample included the system-generated diagnostic \zzh{relative likelihood} score and the physician’s trust rating. Independent two-tailed t-tests (alpha = 0.05) were conducted on both variables.
The results show that the system assigned significantly higher \zzh{relative likelihood} scores to correct diagnoses than to incorrect ones (66.76 ± 14.46 vs. 35.79 ± 16.27; t=6.01, p<0.001, Cohen’s d=2.01). Approximately 90\% of the scores fell within the 20–30 and 60–80 ranges, indicating that \name~ provides meaningful diagnostic separation. Physicians’ trust ratings were also significantly higher for correct diagnoses (81.00 ± 6.56 vs. 29.74 ± 10.78; t=16.98, p<0.001, Cohen’s d=5.67), suggesting that physicians adjusted their trust in accordance with diagnostic accuracy. Overall, within the scope of this study, both system scores and physician trust exhibited consistent and substantial differences across diagnostic correctness, reflecting \name~’s discriminative capability and physicians’ corresponding calibration of trust.

\begin{figure}[t]
    \centering
    \rrfig[width=0.98\linewidth]{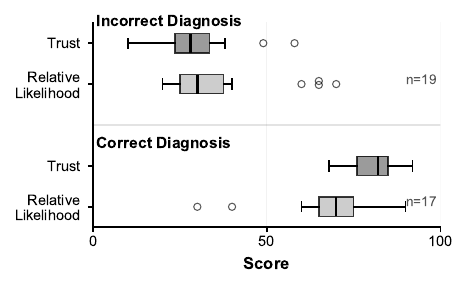} 
    \caption{\rr{\textbf{Diagnostic relative likelihood and trust scores for correct vs. incorrect diagnoses.} Boxplots show system-generated diagnostic relative likelihood scores and physicians' trust ratings (0--100 scale) for cases where the system's diagnosis was correct ($n=17$) or incorrect ($n=19$).}}
    \label{box}
    \Description{The figure contains four horizontal boxplots arranged in two sections labeled Incorrect Diagnosis and Correct Diagnosis. Within each section, the upper boxplot corresponds to Trust and the lower boxplot to Possibility. The x-axis is Score ranging from 0 to 100. In the Incorrect Diagnosis section ($n=19$), the Trust and Possibility boxplots are centered around the middle of the scale, with whiskers extending toward both lower and higher scores and several higher outliers. In the Correct Diagnosis section ($n=17$), both Trust and Possibility boxplots are shifted toward higher scores, with boxes and whiskers mostly between about 60 and 90, indicating generally higher scores than in the incorrect-diagnosis group.}
\end{figure}

}}

\section{\rr{Qualitative Results}}

\rr{This section presents qualitative results of \name~ based on 14 case studies, the user study, and expert interviews. We report three representative case studies and supplementary interview information in Appendix.}

\subsection{\rr{Qualitative Feedback}}
\rr{In this section, we summarize qualitative feedback from the expert interviews.}

\textbf{Usability.}
Experts unanimously agreed that the system is intuitive to use and requires no additional training to get started.
On the one hand, E1 emphasized the value of guided dialogues in medical history collection: \textit{``Patients often do not know how to accurately describe their symptoms,"} noting that the system’s targeted follow-up questions effectively supplement missing information. At the same time, he suggested that after collecting general information, the system should further assist physicians in conducting inquiries to obtain more discriminative clinical details, rather than relying solely on a fixed workflow.
On the other hand, experts expressed appreciation for the diagnostic results and supporting evidence provided by \name~. They highlighted that the clear and structured presentation of diagnoses and evidence enhanced their trust in the system and facilitated a quicker grasp of the diagnostic logic.

\textbf{Willingness to Use.}
Experts reported that in routine diagnostic practice they frequently rely on external tools (\eg, UpToDate) or past case records to obtain relevant medical information. Accordingly, they regarded \name~, which integrates patient cases with a dynamically maintained medical KG, as a system that could substantially reduce their workload.
E2 highlighted the critical importance of accuracy and trustworthiness in clinical practice. He noted: \textit{``I closely follow the progress of AI, but I rarely use GPT in my daily work."}, explaining that he cannot rely on diagnostic results lacking sufficient justification or on medical knowledge generated solely from GPT’s internal parameters. Such concerns, he added, \textit{``may significantly influence physicians’ willingness to adopt AI-driven tools in professional practice."}
By combining the generative capacity of LLMs with a medical KG that is continuously updated and validated by experts, the system enhances the reliability of its outputs and aligns closely with physicians’ evidence-based practices, thereby strengthening their willingness to use it in clinical settings.

\textbf{Visual and Interaction Design.}
E1 strongly endorsed the dynamic presentation of medical history on the patient interface. He remarked: \textit{``In communicating with patients, many barriers arise from their lack of clear understanding of their own condition, which can lead to distrust of physicians."} By gradually disclosing the medical history on the right side of the interface, patients can progressively build a better understanding of their condition. This not only strengthens their trust in both the system and the physician but also facilitates more effective communication regarding diagnosis and follow-up instructions.
E2 emphasized the benefits of automated optimization in diagnostic responses: \textit{``I only need to confirm the factual content of the final diagnosis, and the system automatically generates a polite and appropriate reply. This allows me to focus on professional judgment and reduces the communication burden."} He further suggested integrating automated language refinement into the chat content on the left side of the interface to enhance the system’s scalability.

\textbf{Suggestions for Improvement.}
Experts provided several constructive suggestions for further improving \name~.
E2 recommended more tightly integrating existing knowledge acquisition channels into the evolution of the KG, thereby improving both the quality and efficiency of KG updates while reducing the burden on medical experts. He further suggested that for knowledge points without clear clinical conclusions, the system could incorporate links to relevant medical literature. Although such references may not directly yield diagnostic conclusions, they could guide physicians’ exploration and reduce cognitive load.
E1 emphasized the necessity of enhancing the history-taking function. He noted that collecting patient histories is often time-consuming, effort-intensive, and cognitively demanding. Under the pressures of high-intensity clinical environments, physicians may find it difficult to balance thorough history collection with patients’ emotional needs. Therefore, an interview mode supported by the system not only holds strong practical potential but also offers greater feasibility in real-world clinical practice.

\subsection{\rr{Analysis of Potential Issues.}}

{\rr{

\textbf{Safety.}
\name~ is positioned as a diagnostic aid for physicians rather than an autonomous diagnostic system. Its safety therefore primarily concerns whether its assistance could lead to serious errors in physicians’ judgments. To reduce such risks, \name~ provides, alongside its predicted diagnoses, the estimated diagnostic relative likelihood, the potential severity if each diagnosis were true, structured objective knowledge, and explicit reasoning chains, thereby offering physicians sufficient contextual information. The system’s stand-alone diagnostic performance, human–AI collaborative performance, the usefulness of \zzh{relative likelihood} outputs, and physicians’ trust in these outputs were systematically evaluated in Section \ref{user}, and the results indicate that \name~ is overall effective from a safety perspective. 
To further assess safety, we examined the cases in which, among the 12 scenario packs in Section \ref{user}, the physician assisted by \name~ failed to include a correct diagnosis within their Top-3, specifically the eighth scenario pack in the material\footnote{https://www.thefederation.uk/document/station-2-scenario-pack-16}. In this case, the reference standard diagnosis was nitrofurantoin-induced pulmonary fibrosis, and the acceptable diagnoses included extrinsic allergic alveolitis, chronic obstructive pulmonary disease, and chronic heart failure. Although the system successfully collected the history that the patient had been taking a 50 mg antibiotic every night for a long period as prophylaxis for urinary tract infections, the patient was unable to recall the drug’s name. \name~’s Top-3 preliminary diagnoses were idiopathic pulmonary fibrosis, connective tissue disease–associated interstitial lung disease, and smoking-related interstitial lung disease / combined pulmonary fibrosis and emphysema. After reflection, the physician proposed the same three preliminary diagnoses. According to the hit criteria in this study, this Top-3 set was counted as a miss. After discussing the case with experts (E1–E2), we concluded that, from a clinical safety perspective, both the system and the physician had, at the initial consultation, correctly placed the patient within the spectrum of “chronic progressive fibrotic interstitial lung disease,” rather than misclassifying her as having simple bronchitis or another mild condition. At the same time, \name~ explicitly recommended further investigations, including high-resolution CT, pulmonary function tests with diffusing capacity, and verification of the exact antibiotic name. Once HRCT confirms diffuse fibrotic interstitial changes and prescription records reveal nitrofurantoin use, the physician can narrow the preliminary diagnoses down to the reference standard diagnosis. Thus, although this case did not include the correct diagnosis in the Top-3, the physician, with \name~’s support, was guided along the path toward the correct diagnosis, without deviation or safety concerns. We therefore consider \name~’s safety to be acceptable in this context.

\textbf{History taking.}
We analyzed the issues in the history-taking phase: (1) the patient responded normally, but the extracted history deviated from the template; (2) the patient expressed personal concerns instead of answering; and (3) the patient rejoined after an interruption.
For (1), the system occasionally added non-template fields during template updates (generating a “concomitant symptoms” field when collecting “other medical history”. All similar deviations in the evaluation followed this pattern. This arose from the LLM agent extending fields based on semantic similarity rather than strictly following instructions. Although no information was lost, these deviations increased physicians’ reading burden. We will add an automated consistency check to ensure strict adherence to the predefined structure.
For the second issue, when patients express concerns, the system currently acknowledges them before redirecting the conversation to history taking. Despite improved phrasing through context-aware dialog agent, further refinement is needed. Future versions will enhance emotional awareness so that the agent can first provide emotional support and then resume history taking, allowing multi-turn emotional exchanges without introducing new clinical information. 
For the third issue, the system treats resumed conversations as continuous sessions. We plan to introduce time awareness so that, when interruptions exceed a threshold, the system first re-establishes rapport or reviews the patient’s condition before proceeding.

\textbf{Graph view.}
\name~ provides an objective, disease-centred view of medical knowledge. In terms of presentation, we adopt a conventional knowledge-graph visualisation, using a graph structure that accords with human cognitive habits to help physicians judge whether the system’s diagnostic results are reasonable\cite{yan2024knownet}. However, this widely used display mode has an inherent limitation: when the amount of information is large, it tends to become cluttered and thus reduces reading efficiency. To alleviate this problem, in addition to basic interactions such as dragging and zooming, we rank nodes in the relevant subgraph by importance for the current scenario, and allow physicians to adjust the number of nodes displayed. Nevertheless, we are still concerned that, for diseases with extremely rich associated information, the graph view may become crowded. However, through interviews with multiple physicians and experts, we obtained an counter-intuitive insight: diseases with more associated nodes are often those with which physicians are more familiar, and therefore impose a smaller burden in use. At the same time, among the nodes associated with a disease, the key nodes are usually highly concentrated; even when there are many associated nodes, only a small subset typically plays a core role. On this basis, we argue that, rather than continuing to optimise the graphical presentation itself, a more promising direction is to strengthen the system’s recommendation mechanism. By fully exploiting the reasoning capabilities of large language models, more intelligent recommendations can help physician concentrate their limited cognitive resources on core decision-making tasks.

}}

\section{Discussion}

In this section, we further reflect on the implications of our findings, summarize the broader contributions of \name, and point out the challenges that remain.

\textbf{\rr{Implications} of Collaborative Assisted Diagnosis.}
We proposes \name\, \rr{as an prototype system of a collaborative diagnostic-support paradigm that integrates} LLMs, KGs, and medical experts.
Unlike traditional clinical decision support systems that rely on a single technological pathway, \name\ enables dual interactions between patients and physicians while introducing a \rr{human-in-the-loop} closed-loop reasoning mechanism: LLMs provide semantic understanding and preliminary diagnoses, KGs supply structured medical knowledge, and medical experts \rr{review and calibrate model suggestions, supporting the reliability of diagnostic outcomes and the continuous refinement of knowledge}. This multi-party collaboration \rr{has the potential to} enhance diagnostic accuracy and interpretability \rr{in appropriate use cases} and \rr{offers a concrete example of how digital healthcare tools can be organized around collaborative reasoning rather than fully automated decisions}. Moreover, \rr{the collaborative design of \name\ aligns well with telemedicine workflows and may be adapted to conventional clinical settings.} 
First, in regional medical centers that are under substantial pressure, \name\ \rr{has the potential to} streamline diagnostic workflows, \rr{support} physicians’ efficiency, and \rr{ultimately enhance} patient experiences. Second, in remote clinics with limited access to specialized medical expertise, \name\ \rr{has the potential to enable the} \rr{more} cost-effective introduction of reusable, up-to-date, and \rr{expert-curated} knowledge through expert collaboration.
Thus, our approach \rr{may, in specific deployment contexts, help improve aspects of healthcare equity and narrow certain medical disparities}.

\rr{\textbf{Accessibility and Scope.}
Our evaluation of \name~ was conducted in non-emergency adult internal-medicine consultations using a text-based interface with simulated patients who had at least basic health literacy and did not rely on assistive technologies. 
Consequently, we did not directly assess usability or accessibility for people with low health literacy, limited reading skills, or those who use screen readers or alternative input devices, and our findings may overestimate its suitability for these groups. 
Although the underlying knowledge graph covers a broad range of conditions, the current instantiation and case set focus on common internal-medicine presentations. 
We position \name~ as a physician-in-the-loop decision-support tool for relatively stable, non-surgical encounters in outpatient-like settings, rather than as a general-purpose autonomous diagnostic system, and we emphasise that this work presents a prototype system. 
Substantial additional research, including rigorous evaluation of safety, reliability, effectiveness, and privacy, will be required before \name~ can be safely used by healthcare providers, administrators, or patients. 
Future work will extend the patient-facing interface with multimodal, accessibility-oriented interaction (e.g., speech input and output, simplified and layered explanations, optional digital-human agents) and will co-design and evaluate these variants with patients who have diverse communication abilities and technology needs, while extending the system to a broader set of clinical domains.}

\rr{\textbf{Limitations and Future Work.}
Our efforts in improving the diagnostic experience of doctors and patients with Diaglink unveil three future research opportunities.
First, there is a clear asymmetry of needs: patients care about consultation experience, while the burden of linguistic and emotional communication reduces physicians’ work satisfaction and increases pressure on limited medical resources. The constrained LLM introduced in this study shows preliminary effects in mitigating this asymmetry. Future work may explore dynamically adjustable, more targeted case templates, combined with reinforcement learning, to further improve its interaction and instruction-following capabilities in this setting. 
Second, in this vertical task, physicians generally welcome the system’s proactive recommendation of relevant information. In contrast to relatively mature physician-side retrieval tools and recent multi-agent retrieval methods, this recommendation remains less developed; multi-agent–based recommendation therefore 
has the potential to become a more effective approach.
Third, our evaluation contains limitations: the effect of KG evolution requires long-term observation in real deployment, which is beyond this prototype study, and experimental patients were played by trained volunteers, who cannot fully reproduce the complexity of real cases. In the future, we hope to conduct long-term, real-world user studies under appropriate safety and ethics approvals.}

\section{Conclusion}

We propose \name, a novel dual-user interactive diagnostic support system that integrates LLMs, KGs, and medical experts into a closed-loop collaborative framework.
Through guided dialogue, \name\ enables patients to provide accurate and comprehensive medical histories with reduced patients’ communication burden. Its collaborative diagnostic module combines the contextual reasoning capabilities of LLMs with the structured reliability of KGs to generate accurate and interpretable diagnostic recommendations. The “physician-in-the-loop” mechanism allows medical experts to validate, refine, and continuously evolve the KG, ensuring that diagnostic knowledge remains up-to-date and clinically relevant over time.
We demonstrate the effectiveness of \name\ in supporting diagnostic decision-making through \rr{user study,} use cases and expert interviews. For future work, we plan to extend the system to support direct physician–patient interactions and incorporate multimodal data sources, further enhancing \name’s  capabilities.

\begin{acks}
The work was partially supported by the National Key Research and Development Program of China (No. 2024YFF0617702); the National Natural Science Foundation of China (Nos. U22A2025, 62402097, 62232007, and U23A20309); the Joint Funds of the Natural Science Foundation of Liaoning Province (No. 2023-BSBA-132); the 111 Project (No. B16009); the Ant Group Research Program (No. 2025021900003); and the Fundamental Research Funds for the Central Universities (No. N2417007).
\end{acks}
\bibliographystyle{ACM-Reference-Format}
\bibliography{2_CHI-2026}


\begin{thebibliography}{104}


\ifx \showCODEN    \undefined \def \showCODEN     #1{\unskip}     \fi
\ifx \showISBNx    \undefined \def \showISBNx     #1{\unskip}     \fi
\ifx \showISBNxiii \undefined \def \showISBNxiii  #1{\unskip}     \fi
\ifx \showISSN     \undefined \def \showISSN      #1{\unskip}     \fi
\ifx \showLCCN     \undefined \def \showLCCN      #1{\unskip}     \fi
\ifx \shownote     \undefined \def \shownote      #1{#1}          \fi
\ifx \showarticletitle \undefined \def \showarticletitle #1{#1}   \fi
\ifx \showURL      \undefined \def \showURL       {\relax}        \fi
\providecommand\bibfield[2]{#2}
\providecommand\bibinfo[2]{#2}
\providecommand\natexlab[1]{#1}
\providecommand\showeprint[2][]{arXiv:#2}

\bibitem[nux(2025)]%
        {nuxt}
 \bibinfo{year}{2025}\natexlab{}.
\newblock \bibinfo{title}{Nuxt: The Progressive Web Framework}.
\newblock \bibinfo{howpublished}{\url{https://nuxt.com/}}.
\newblock
\newblock
\shownote{Nuxt official website, accessed 2025-12-04}.


\bibitem[{Ada Health GmbH}(2025)]%
        {AdaHealth2025}
\bibfield{author}{\bibinfo{person}{{Ada Health GmbH}}.} \bibinfo{year}{2025}\natexlab{}.
\newblock \bibinfo{title}{Ada: AI-powered Symptom Checker}.
\newblock \bibinfo{howpublished}{\url{https://ada.com/app}}.
\newblock


\bibitem[Adams et~al\mbox{.}(2022)]%
        {adams2022prospective}
\bibfield{author}{\bibinfo{person}{Roy Adams}, \bibinfo{person}{Katharine~E Henry}, \bibinfo{person}{Anirudh Sridharan}, \bibinfo{person}{Hossein Soleimani}, \bibinfo{person}{Andong Zhan}, \bibinfo{person}{Nishi Rawat}, \bibinfo{person}{Lauren Johnson}, \bibinfo{person}{David~N Hager}, \bibinfo{person}{Sara~E Cosgrove}, \bibinfo{person}{Andrew Markowski}, {et~al\mbox{.}}} \bibinfo{year}{2022}\natexlab{}.
\newblock \showarticletitle{Prospective, multi-site study of patient outcomes after implementation of the TREWS machine learning-based early warning system for sepsis}.
\newblock \bibinfo{journal}{\emph{Nature medicine}} \bibinfo{volume}{28}, \bibinfo{number}{7} (\bibinfo{year}{2022}), \bibinfo{pages}{1455--1460}.
\newblock
\href{https://doi.org/10.1038/s41591-022-01894-0}{doi:\nolinkurl{10.1038/s41591-022-01894-0}}


\bibitem[Amershi et~al\mbox{.}(2014)]%
        {Amershi2014Power}
\bibfield{author}{\bibinfo{person}{Saleema Amershi}, \bibinfo{person}{Maya Cakmak}, \bibinfo{person}{W.~Bradley Knox}, {and} \bibinfo{person}{Todd Kulesza}.} \bibinfo{year}{2014}\natexlab{}.
\newblock \showarticletitle{Power to the People: The Role of Humans in Interactive Machine Learning}.
\newblock \bibinfo{journal}{\emph{AI Magazine}} \bibinfo{volume}{35}, \bibinfo{number}{4} (\bibinfo{year}{2014}), \bibinfo{pages}{105--120}.
\newblock
\href{https://doi.org/10.1609/aimag.v35i4.2513}{doi:\nolinkurl{10.1609/aimag.v35i4.2513}}


\bibitem[Amershi et~al\mbox{.}(2019)]%
        {Amershi2019Guidelines}
\bibfield{author}{\bibinfo{person}{Saleema Amershi}, \bibinfo{person}{Dan Weld}, \bibinfo{person}{Mihaela Vorvoreanu}, \bibinfo{person}{Adam Fourney}, \bibinfo{person}{Besmira Nushi}, \bibinfo{person}{Penny Collisson}, \bibinfo{person}{Jina Suh}, \bibinfo{person}{Shamsi Iqbal}, \bibinfo{person}{Paul~N. Bennett}, \bibinfo{person}{Kori Inkpen}, \bibinfo{person}{Jaime Teevan}, \bibinfo{person}{Ruth Kikin{-}Gil}, {and} \bibinfo{person}{Eric Horvitz}.} \bibinfo{year}{2019}\natexlab{}.
\newblock \showarticletitle{Guidelines for Human-{AI} Interaction}. In \bibinfo{booktitle}{\emph{Proceedings of the 2019 {CHI} Conference on Human Factors in Computing Systems}} \emph{(\bibinfo{series}{CHI '19})}. \bibinfo{publisher}{Association for Computing Machinery}, \bibinfo{address}{Glasgow, Scotland, UK}, Article \bibinfo{articleno}{3}, \bibinfo{numpages}{13}~pages.
\newblock
\href{https://doi.org/10.1145/3290605.3300233}{doi:\nolinkurl{10.1145/3290605.3300233}}


\bibitem[Asgari et~al\mbox{.}(2024)]%
        {asgari2024_ehr_cognitive_load}
\bibfield{author}{\bibinfo{person}{Elham Asgari}, \bibinfo{person}{Japsimar Kaur}, \bibinfo{person}{Gani Nuredini}, \bibinfo{person}{Jasmine Balloch}, \bibinfo{person}{Andrew~M. Taylor}, \bibinfo{person}{Neil Sebire}, \bibinfo{person}{Robert Robinson}, \bibinfo{person}{Catherine Peters}, \bibinfo{person}{Shankar Sridharan}, {and} \bibinfo{person}{Dominic Pimenta}.} \bibinfo{year}{2024}\natexlab{}.
\newblock \showarticletitle{Impact of Electronic Health Record Use on Cognitive Load and Burnout Among Clinicians: Narrative Review}.
\newblock \bibinfo{journal}{\emph{JMIR Medical Informatics}}  \bibinfo{volume}{12} (\bibinfo{year}{2024}), \bibinfo{pages}{e55499}.
\newblock
\href{https://doi.org/10.2196/55499}{doi:\nolinkurl{10.2196/55499}}


\bibitem[Ayers et~al\mbox{.}(2023)]%
        {Ayers2023JAMAIM}
\bibfield{author}{\bibinfo{person}{John~W. Ayers}, \bibinfo{person}{Adam Poliak}, \bibinfo{person}{Mark Dredze}, \bibinfo{person}{Eric~C. Leas}, \bibinfo{person}{Zechariah Zhu}, \bibinfo{person}{Jessica~B. Kelley}, \bibinfo{person}{Dennis~J. Faix}, \bibinfo{person}{Aaron~M. Goodman}, \bibinfo{person}{Christopher~A. Longhurst}, \bibinfo{person}{Michael Hogarth}, {and} \bibinfo{person}{Davey~M. Smith}.} \bibinfo{year}{2023}\natexlab{}.
\newblock \showarticletitle{Comparing Physician and Artificial Intelligence Chatbot Responses to Patient Questions Posted to a Public Social Media Forum}.
\newblock \bibinfo{journal}{\emph{JAMA Internal Medicine}} \bibinfo{volume}{183}, \bibinfo{number}{6} (\bibinfo{year}{2023}), \bibinfo{pages}{589--596}.
\newblock
\href{https://doi.org/10.1001/jamainternmed.2023.1838}{doi:\nolinkurl{10.1001/jamainternmed.2023.1838}}


\bibitem[Baek et~al\mbox{.}(2023)]%
        {baek2023knowledge}
\bibfield{author}{\bibinfo{person}{Jinheon Baek}, \bibinfo{person}{Soyeong Jeong}, \bibinfo{person}{Minki Kang}, \bibinfo{person}{Jong~C Park}, {and} \bibinfo{person}{Sung~Ju Hwang}.} \bibinfo{year}{2023}\natexlab{}.
\newblock \showarticletitle{Knowledge-Augmented Language Model Verification}. In \bibinfo{booktitle}{\emph{Proceedings of the 2023 Conference on Empirical Methods in Natural Language Processing}} (Singapore). \bibinfo{publisher}{Association for Computational Linguistics}, \bibinfo{address}{Minneapolis, Minnesota}, \bibinfo{pages}{1720--1736}.
\newblock
\href{https://doi.org/10.18653/v1/2023.emnlp-main.107}{doi:\nolinkurl{10.18653/v1/2023.emnlp-main.107}}


\bibitem[Bangor et~al\mbox{.}(2008)]%
        {Bangor2008SUS}
\bibfield{author}{\bibinfo{person}{Aaron Bangor}, \bibinfo{person}{Philip~T. Kortum}, {and} \bibinfo{person}{James~T. Miller}.} \bibinfo{year}{2008}\natexlab{}.
\newblock \showarticletitle{An Empirical Evaluation of the System Usability Scale}.
\newblock \bibinfo{journal}{\emph{International Journal of Human–Computer Interaction}} \bibinfo{volume}{24}, \bibinfo{number}{6} (\bibinfo{year}{2008}), \bibinfo{pages}{574--594}.
\newblock
\href{https://doi.org/10.1080/10447310802205776}{doi:\nolinkurl{10.1080/10447310802205776}}


\bibitem[Bleher and Braun(2022)]%
        {bleher2022_diffused_responsibility}
\bibfield{author}{\bibinfo{person}{Hannah Bleher} {and} \bibinfo{person}{Matthias Braun}.} \bibinfo{year}{2022}\natexlab{}.
\newblock \showarticletitle{Diffused Responsibility: Attributions of Responsibility in the Use of {AI}-Driven Clinical Decision Support Systems}.
\newblock \bibinfo{journal}{\emph{AI and Ethics}} \bibinfo{volume}{2}, \bibinfo{number}{4} (\bibinfo{year}{2022}), \bibinfo{pages}{747--761}.
\newblock
\href{https://doi.org/10.1007/s43681-022-00135-x}{doi:\nolinkurl{10.1007/s43681-022-00135-x}}


\bibitem[Bo et~al\mbox{.}(2024)]%
        {bo2024incremental}
\bibfield{author}{\bibinfo{person}{Jessica~Y Bo}, \bibinfo{person}{Pan Hao}, {and} \bibinfo{person}{Brian~Y Lim}.} \bibinfo{year}{2024}\natexlab{}.
\newblock \showarticletitle{Incremental xai: Memorable understanding of ai with incremental explanations}. In \bibinfo{booktitle}{\emph{Proceedings of the 2024 CHI Conference on Human Factors in Computing Systems}} (Honolulu, HI, USA). \bibinfo{publisher}{ACM}, \bibinfo{address}{New York, NY, USA}, \bibinfo{pages}{1--17}.
\newblock
\href{https://doi.org/10.1145/3613904.3642689}{doi:\nolinkurl{10.1145/3613904.3642689}}


\bibitem[Brooke(1996)]%
        {Brooke1996SUS}
\bibfield{author}{\bibinfo{person}{John Brooke}.} \bibinfo{year}{1996}\natexlab{}.
\newblock \showarticletitle{SUS: A ``quick and dirty'' usability scale}.
\newblock In \bibinfo{booktitle}{\emph{Usability Evaluation in Industry}}, \bibfield{editor}{\bibinfo{person}{Patrick~W. Jordan}, \bibinfo{person}{Bruce Thomas}, \bibinfo{person}{Ian~L. Weerdmeester}, {and} \bibinfo{person}{A.~L. McClelland}} (Eds.). \bibinfo{publisher}{Taylor \& Francis}, \bibinfo{address}{London}, Chapter~21, \bibinfo{pages}{189--194}.
\newblock
\showISBNx{978-0-7484-0486-3}


\bibitem[Chandak et~al\mbox{.}(2023)]%
        {Chandak2023PrimeKG}
\bibfield{author}{\bibinfo{person}{Payal Chandak}, \bibinfo{person}{Kexin Huang}, {and} \bibinfo{person}{Marinka Zitnik}.} \bibinfo{year}{2023}\natexlab{}.
\newblock \showarticletitle{Building a knowledge graph to enable precision medicine}.
\newblock \bibinfo{journal}{\emph{Scientific Data}}  \bibinfo{volume}{10} (\bibinfo{year}{2023}), \bibinfo{pages}{67}.
\newblock
\href{https://doi.org/10.1038/s41597-023-01960-3}{doi:\nolinkurl{10.1038/s41597-023-01960-3}}


\bibitem[Devarakonda et~al\mbox{.}(2024)]%
        {Devarakonda2024CTKG}
\bibfield{author}{\bibinfo{person}{Murthy~V. Devarakonda}, \bibinfo{person}{Smita Mohanty}, \bibinfo{person}{Raja~Rao Sunkishala}, \bibinfo{person}{Nag Mallampalli}, {and} \bibinfo{person}{Xiong Liu}.} \bibinfo{year}{2024}\natexlab{}.
\newblock \showarticletitle{Clinical trial recommendations using Semantics-Based inductive inference and knowledge graph embeddings}.
\newblock \bibinfo{journal}{\emph{J. of Biomedical Informatics}} \bibinfo{volume}{154}, \bibinfo{number}{C} (\bibinfo{date}{June} \bibinfo{year}{2024}), \bibinfo{numpages}{12}~pages.
\newblock
\showISSN{1532-0464}
\href{https://doi.org/10.1016/j.jbi.2024.104627}{doi:\nolinkurl{10.1016/j.jbi.2024.104627}}


\bibitem[Dourish(2006)]%
        {Dourish2006Implications}
\bibfield{author}{\bibinfo{person}{Paul Dourish}.} \bibinfo{year}{2006}\natexlab{}.
\newblock \showarticletitle{Implications for Design}. In \bibinfo{booktitle}{\emph{Proceedings of the {SIGCHI} Conference on Human Factors in Computing Systems}} \emph{(\bibinfo{series}{CHI '06})}. \bibinfo{publisher}{Association for Computing Machinery}, \bibinfo{address}{Montr{\'e}al, Qu{\'e}bec, Canada}, \bibinfo{pages}{541--550}.
\newblock
\href{https://doi.org/10.1145/1124772.1124855}{doi:\nolinkurl{10.1145/1124772.1124855}}


\bibitem[Douze et~al\mbox{.}(2024)]%
        {douze2024faiss}
\bibfield{author}{\bibinfo{person}{Matthijs Douze}, \bibinfo{person}{Alexandr Guzhva}, \bibinfo{person}{Chengqi Deng}, \bibinfo{person}{Jeff Johnson}, \bibinfo{person}{Gergely Szilvasy}, \bibinfo{person}{Pierre-Emmanuel Mazar{\'e}}, \bibinfo{person}{Maria Lomeli}, \bibinfo{person}{Lucas Hosseini}, {and} \bibinfo{person}{Herv{\'e} J{\'e}gou}.} \bibinfo{year}{2024}\natexlab{}.
\newblock \bibinfo{title}{The faiss library}.
\newblock
\showeprint[arxiv]{2401.08281}~[cs.LG]
\href{https://doi.org/10.48550/arXiv.2401.08281}{doi:\nolinkurl{10.48550/arXiv.2401.08281}}


\bibitem[Fahrner et~al\mbox{.}(2025)]%
        {fahrner2025generative}
\bibfield{author}{\bibinfo{person}{L~John Fahrner}, \bibinfo{person}{Emma Chen}, \bibinfo{person}{Eric Topol}, {and} \bibinfo{person}{Pranav Rajpurkar}.} \bibinfo{year}{2025}\natexlab{}.
\newblock \showarticletitle{The generative era of medical AI}.
\newblock \bibinfo{journal}{\emph{Cell}} \bibinfo{volume}{188}, \bibinfo{number}{14} (\bibinfo{year}{2025}), \bibinfo{pages}{3648--3660}.
\newblock
\href{https://doi.org/10.1016/j.cell.2025.05.018}{doi:\nolinkurl{10.1016/j.cell.2025.05.018}}


\bibitem[Fan et~al\mbox{.}(2025)]%
        {fan2024visual}
\bibfield{author}{\bibinfo{person}{Mengjie Fan}, \bibinfo{person}{Jinlu Yu}, \bibinfo{person}{Daniel Weiskopf}, \bibinfo{person}{Nan Cao}, \bibinfo{person}{Huai-Yu Wang}, {and} \bibinfo{person}{Liang Zhou}.} \bibinfo{year}{2025}\natexlab{}.
\newblock \showarticletitle{Visual analysis of multi-outcome causal graphs}.
\newblock \bibinfo{journal}{\emph{IEEE Transactions on Visualization and Computer Graphics}} \bibinfo{volume}{31}, \bibinfo{number}{1} (\bibinfo{year}{2025}), \bibinfo{pages}{656--666}.
\newblock
\href{https://doi.org/10.1109/TVCG.2024.3456346}{doi:\nolinkurl{10.1109/TVCG.2024.3456346}}


\bibitem[Feng et~al\mbox{.}(2024a)]%
        {feng2024trafps}
\bibfield{author}{\bibinfo{person}{Zezheng Feng}, \bibinfo{person}{Yifan Jiang}, \bibinfo{person}{Hongjun Wang}, \bibinfo{person}{Zipei Fan}, \bibinfo{person}{Yuxin Ma}, \bibinfo{person}{Shuang-Hua Yang}, \bibinfo{person}{Huamin Qu}, {and} \bibinfo{person}{Xuan Song}.} \bibinfo{year}{2024}\natexlab{a}.
\newblock \showarticletitle{TrafPS: A {Shapley}-Based Visual Analytics Approach to Interpret Traffic}.
\newblock \bibinfo{journal}{\emph{Computational Visual Media}} \bibinfo{volume}{10}, \bibinfo{number}{6} (\bibinfo{year}{2024}), \bibinfo{pages}{1101--1119}.
\newblock


\bibitem[Feng et~al\mbox{.}(2020)]%
        {feng2020topology}
\bibfield{author}{\bibinfo{person}{Zezheng Feng}, \bibinfo{person}{Haotian Li}, \bibinfo{person}{Wei Zeng}, \bibinfo{person}{Shuang-Hua Yang}, {and} \bibinfo{person}{Huamin Qu}.} \bibinfo{year}{2020}\natexlab{}.
\newblock \showarticletitle{Topology Density Map for Urban Data Visualization and Analysis}.
\newblock \bibinfo{journal}{\emph{IEEE Transactions on Visualization and Computer Graphics}} \bibinfo{volume}{27}, \bibinfo{number}{2} (\bibinfo{year}{2020}), \bibinfo{pages}{828--838}.
\newblock


\bibitem[Feng et~al\mbox{.}(2022)]%
        {feng2022survey}
\bibfield{author}{\bibinfo{person}{Zezheng Feng}, \bibinfo{person}{Huamin Qu}, \bibinfo{person}{Shuang-Hua Yang}, \bibinfo{person}{Yulong Ding}, {and} \bibinfo{person}{Jie Song}.} \bibinfo{year}{2022}\natexlab{}.
\newblock \showarticletitle{A Survey of Visual Analytics in Urban Area}.
\newblock \bibinfo{journal}{\emph{Expert Systems}} \bibinfo{volume}{39}, \bibinfo{number}{9} (\bibinfo{year}{2022}), \bibinfo{pages}{e13065}.
\newblock


\bibitem[Feng et~al\mbox{.}(2024b)]%
        {feng2024holens}
\bibfield{author}{\bibinfo{person}{Zezheng Feng}, \bibinfo{person}{Fang Zhu}, \bibinfo{person}{Hongjun Wang}, \bibinfo{person}{Jianing Hao}, \bibinfo{person}{Shuang-Hua Yang}, \bibinfo{person}{Wei Zeng}, {and} \bibinfo{person}{Huamin Qu}.} \bibinfo{year}{2024}\natexlab{b}.
\newblock \showarticletitle{{HoLens}: A Visual Analytics Design for Higher-Order Movement Modeling and Visualization}.
\newblock \bibinfo{journal}{\emph{Computational Visual Media}} \bibinfo{volume}{10}, \bibinfo{number}{6} (\bibinfo{year}{2024}), \bibinfo{pages}{1079--1100}.
\newblock


\bibitem[Fleming et~al\mbox{.}(2021)]%
        {Fleming2021Diagnostics}
\bibfield{author}{\bibinfo{person}{Kenneth~A. Fleming}, \bibinfo{person}{Susan Horton}, \bibinfo{person}{Michael~L. Wilson}, \bibinfo{person}{Rifat Atun}, \bibinfo{person}{Kristen DeStigter}, \bibinfo{person}{John Flanigan}, \bibinfo{person}{Shahin Sayed}, \bibinfo{person}{Pierrick Adam}, \bibinfo{person}{Bertha Aguilar}, \bibinfo{person}{Savvas Andronikou}, \bibinfo{person}{Catharina Boehme}, \bibinfo{person}{William Cherniak}, \bibinfo{person}{Annie N.~Y. Cheung}, \bibinfo{person}{Bernice Dahn}, \bibinfo{person}{Lluis Donoso-Bach}, \bibinfo{person}{Tania Douglas}, \bibinfo{person}{Patricia Garcia}, \bibinfo{person}{Sarwat Hussain}, \bibinfo{person}{Hari~S. Iyer}, \bibinfo{person}{Mikashmi Kohli}, \bibinfo{person}{Alain~B. Labrique}, \bibinfo{person}{Lai~Meng Looi}, \bibinfo{person}{John~G. Meara}, \bibinfo{person}{John Nkengasong}, \bibinfo{person}{Madhukar Pai}, \bibinfo{person}{Kara~Lee Pool}, \bibinfo{person}{Kaushik Ramaiya}, \bibinfo{person}{Lee Schroeder}, \bibinfo{person}{Devanshi Shah}, \bibinfo{person}{Richard Sullivan}, \bibinfo{person}{Bien~Soo Tan}, {and} \bibinfo{person}{Kamini Walia}.} \bibinfo{year}{2021}\natexlab{}.
\newblock \showarticletitle{The Lancet Commission on diagnostics: transforming access to diagnostics}.
\newblock \bibinfo{journal}{\emph{The Lancet}} \bibinfo{volume}{398}, \bibinfo{number}{10315} (\bibinfo{date}{Nov} \bibinfo{year}{2021}), \bibinfo{pages}{1997--2050}.
\newblock
\href{https://doi.org/10.1016/S0140-6736(21)00673-5}{doi:\nolinkurl{10.1016/S0140-6736(21)00673-5}}


\bibitem[Fraser et~al\mbox{.}(2023)]%
        {fraser2023comparison}
\bibfield{author}{\bibinfo{person}{Hamish Fraser}, \bibinfo{person}{Daven Crossland}, \bibinfo{person}{Ian Bacher}, \bibinfo{person}{Megan Ranney}, \bibinfo{person}{Tracy Madsen}, \bibinfo{person}{Ross Hilliard}, {et~al\mbox{.}}} \bibinfo{year}{2023}\natexlab{}.
\newblock \showarticletitle{Comparison of diagnostic and triage accuracy of Ada health and WebMD symptom checkers, ChatGPT, and physicians for patients in an emergency department: clinical data analysis study}.
\newblock \bibinfo{journal}{\emph{JMIR mHealth and uHealth}} \bibinfo{volume}{11}, \bibinfo{number}{1}, Article \bibinfo{articleno}{e49995} (\bibinfo{year}{2023}), \bibinfo{numpages}{13}~pages.
\newblock
\href{https://doi.org/10.2196/49995}{doi:\nolinkurl{10.2196/49995}}


\bibitem[Gao et~al\mbox{.}(2025)]%
        {gao2025healthgenie}
\bibfield{author}{\bibinfo{person}{Fan Gao}, \bibinfo{person}{Xinjie Zhao}, \bibinfo{person}{Ding Xia}, \bibinfo{person}{Zhongyi Zhou}, \bibinfo{person}{Rui Yang}, \bibinfo{person}{Jinghui Lu}, \bibinfo{person}{Hang Jiang}, \bibinfo{person}{Chanjun Park}, {and} \bibinfo{person}{Irene Li}.} \bibinfo{year}{2025}\natexlab{}.
\newblock \bibinfo{title}{HealthGenie: Empowering Users with Healthy Dietary Guidance through Knowledge Graph and Large Language Models}.
\newblock
\showeprint[arxiv]{2504.14594}~[cs.AI]
\href{https://doi.org/10.48550/arXiv.2504.14594}{doi:\nolinkurl{10.48550/arXiv.2504.14594}}


\bibitem[Ge et~al\mbox{.}(2025)]%
        {Ge2025Gated}
\bibfield{author}{\bibinfo{person}{Zhaodi Ge}, \bibinfo{person}{Hanning Chen}, \bibinfo{person}{Xiaodan Liang}, {and} \bibinfo{person}{Lianbo Ma}.} \bibinfo{year}{2025}\natexlab{}.
\newblock \showarticletitle{Gated Mechanism Attention Transformer Based on Wavelet Enhanced Optical Flow Field Estimation for Foreground Detection}.
\newblock \bibinfo{journal}{\emph{IEEE Transactions on Circuits and Systems for Video Technology}} \bibinfo{volume}{35}, \bibinfo{number}{4} (\bibinfo{date}{apr} \bibinfo{year}{2025}).
\newblock
\href{https://doi.org/10.1109/TCSVT.2024.3509952}{doi:\nolinkurl{10.1109/TCSVT.2024.3509952}}


\bibitem[Greenberg and Buxton(2008)]%
        {Greenberg2008Usability}
\bibfield{author}{\bibinfo{person}{Saul Greenberg} {and} \bibinfo{person}{William Buxton}.} \bibinfo{year}{2008}\natexlab{}.
\newblock \showarticletitle{Usability Evaluation Considered Harmful (Some of the Time)}. In \bibinfo{booktitle}{\emph{Proceedings of the 2008 {CHI} Conference on Human Factors in Computing Systems}} \emph{(\bibinfo{series}{CHI '08})}. \bibinfo{publisher}{Association for Computing Machinery}, \bibinfo{address}{Florence, Italy}, \bibinfo{pages}{111--120}.
\newblock
\href{https://doi.org/10.1145/1357054.1357074}{doi:\nolinkurl{10.1145/1357054.1357074}}


\bibitem[Hao et~al\mbox{.}(2023)]%
        {hao2023hofd}
\bibfield{author}{\bibinfo{person}{Shuang Hao}, \bibinfo{person}{Chengliang Chai}, \bibinfo{person}{Guoliang Li}, \bibinfo{person}{Nan Tang}, \bibinfo{person}{Ning Wang}, {and} \bibinfo{person}{Xiang Yu}.} \bibinfo{year}{2023}\natexlab{}.
\newblock \showarticletitle{HOFD: An outdated fact detector for knowledge bases}.
\newblock \bibinfo{journal}{\emph{IEEE Transactions on Knowledge and Data Engineering}} \bibinfo{volume}{35}, \bibinfo{number}{10} (\bibinfo{year}{2023}), \bibinfo{pages}{10775--10789}.
\newblock
\href{https://doi.org/10.1109/TKDE.2023.3248223}{doi:\nolinkurl{10.1109/TKDE.2023.3248223}}


\bibitem[Harnoune et~al\mbox{.}(2021)]%
        {harnoune2021bert}
\bibfield{author}{\bibinfo{person}{Ayoub Harnoune}, \bibinfo{person}{Maryem Rhanoui}, \bibinfo{person}{Mounia Mikram}, \bibinfo{person}{Siham Yousfi}, \bibinfo{person}{Zineb Elkaimbillah}, {and} \bibinfo{person}{Bouchra El~Asri}.} \bibinfo{year}{2021}\natexlab{}.
\newblock \showarticletitle{BERT based clinical knowledge extraction for biomedical knowledge graph construction and analysis}.
\newblock \bibinfo{journal}{\emph{Computer Methods and Programs in Biomedicine Update}}  \bibinfo{volume}{1}, Article \bibinfo{articleno}{100042} (\bibinfo{year}{2021}), \bibinfo{numpages}{7}~pages.
\newblock
\href{https://doi.org/10.1016/j.cmpbup.2021.100042}{doi:\nolinkurl{10.1016/j.cmpbup.2021.100042}}


\bibitem[Hart and Staveland(1988)]%
        {Hart1988NASA_TLX}
\bibfield{author}{\bibinfo{person}{Sandra~G. Hart} {and} \bibinfo{person}{Lowell~E. Staveland}.} \bibinfo{year}{1988}\natexlab{}.
\newblock \showarticletitle{Development of NASA-TLX (Task Load Index): Results of Empirical and Theoretical Research}.
\newblock In \bibinfo{booktitle}{\emph{Advances in Psychology}}, \bibfield{editor}{\bibinfo{person}{Peter~A. Hancock} {and} \bibinfo{person}{Najmedin Meshkati}} (Eds.). Vol.~\bibinfo{volume}{52}. \bibinfo{publisher}{North-Holland}, \bibinfo{address}{Amsterdam}, Chapter~7, \bibinfo{pages}{139--183}.
\newblock
\showISBNx{978-0-444-70388-0}
\href{https://doi.org/10.1016/S0166-4115(08)62386-9}{doi:\nolinkurl{10.1016/S0166-4115(08)62386-9}}


\bibitem[He et~al\mbox{.}(2016)]%
        {he2016deep}
\bibfield{author}{\bibinfo{person}{Kaiming He}, \bibinfo{person}{Xiangyu Zhang}, \bibinfo{person}{Shaoqing Ren}, {and} \bibinfo{person}{Jian Sun}.} \bibinfo{year}{2016}\natexlab{}.
\newblock \showarticletitle{Deep residual learning for image recognition}. In \bibinfo{booktitle}{\emph{Proceedings of the IEEE conference on computer vision and pattern recognition}}. \bibinfo{publisher}{IEEE Computer Society}, \bibinfo{address}{Los Alamitos, CA, USA}, \bibinfo{pages}{770--778}.
\newblock
\href{https://doi.org/10.1109/CVPR.2016.90}{doi:\nolinkurl{10.1109/CVPR.2016.90}}


\bibitem[Hricak et~al\mbox{.}(2021)]%
        {Hricak2021Imaging}
\bibfield{author}{\bibinfo{person}{Hedvig Hricak}, \bibinfo{person}{May Abdel‑Wahab}, \bibinfo{person}{Rifat Atun}, \bibinfo{person}{Miriam~Mikhail Lette}, \bibinfo{person}{Diana Paez}, \bibinfo{person}{James~A. Brink}, \bibinfo{person}{Lluis Donoso‑Bach}, \bibinfo{person}{Guy Frija}, \bibinfo{person}{Monika Hierath}, \bibinfo{person}{Ola Holmberg}, \bibinfo{person}{Pek‑Lan Khong}, \bibinfo{person}{Jason~S. Lewis}, \bibinfo{person}{Geraldine McGinty}, \bibinfo{person}{Wim J.~G. Oyen}, \bibinfo{person}{Lawrence~N. Shulman}, \bibinfo{person}{Zachary~J. Ward}, {and} \bibinfo{person}{Andrew~M. Scott}.} \bibinfo{year}{2021}\natexlab{}.
\newblock \showarticletitle{Medical imaging and nuclear medicine: a Lancet Oncology Commission}.
\newblock \bibinfo{journal}{\emph{The Lancet Oncology}} \bibinfo{volume}{22}, \bibinfo{number}{4} (\bibinfo{date}{Apr} \bibinfo{year}{2021}), \bibinfo{pages}{e136--e172}.
\newblock
\href{https://doi.org/10.1016/S1470-2045(20)30751-8}{doi:\nolinkurl{10.1016/S1470-2045(20)30751-8}}


\bibitem[Hsuan~Yuan et~al\mbox{.}(2024)]%
        {yuan2024kgscope}
\bibfield{author}{\bibinfo{person}{Chao-Wen Hsuan~Yuan}, \bibinfo{person}{Tzu-Wei Yu}, \bibinfo{person}{Jia-Yu Pan}, {and} \bibinfo{person}{Wen-Chieh Lin}.} \bibinfo{year}{2024}\natexlab{}.
\newblock \showarticletitle{KGScope: Interactive Visual Exploration of Knowledge Graphs With Embedding-Based Guidance}.
\newblock \bibinfo{journal}{\emph{IEEE Transactions on Visualization and Computer Graphics}} \bibinfo{volume}{30}, \bibinfo{number}{12} (\bibinfo{year}{2024}), \bibinfo{pages}{7702--7716}.
\newblock
\href{https://doi.org/10.1109/TVCG.2024.3360690}{doi:\nolinkurl{10.1109/TVCG.2024.3360690}}


\bibitem[Huang et~al\mbox{.}(2025)]%
        {huang2025survey}
\bibfield{author}{\bibinfo{person}{Lei Huang}, \bibinfo{person}{Weijiang Yu}, \bibinfo{person}{Weitao Ma}, \bibinfo{person}{Weihong Zhong}, \bibinfo{person}{Zhangyin Feng}, \bibinfo{person}{Haotian Wang}, \bibinfo{person}{Qianglong Chen}, \bibinfo{person}{Weihua Peng}, \bibinfo{person}{Xiaocheng Feng}, \bibinfo{person}{Bing Qin}, {et~al\mbox{.}}} \bibinfo{year}{2025}\natexlab{}.
\newblock \showarticletitle{A survey on hallucination in large language models: Principles, taxonomy, challenges, and open questions}.
\newblock \bibinfo{journal}{\emph{ACM Transactions on Information Systems}} \bibinfo{volume}{43}, \bibinfo{number}{2} (\bibinfo{year}{2025}), \bibinfo{pages}{1--55}.
\newblock
\href{https://doi.org/10.1145/3703155}{doi:\nolinkurl{10.1145/3703155}}


\bibitem[Husain et~al\mbox{.}(2021)]%
        {husain2021multi}
\bibfield{author}{\bibinfo{person}{Fahd Husain}, \bibinfo{person}{Rosa Romero-Gomez}, \bibinfo{person}{Emily Kuang}, \bibinfo{person}{Dario Segura}, \bibinfo{person}{Adamo Carolli}, \bibinfo{person}{Lai~Chung Liu}, \bibinfo{person}{Manfred Cheung}, {and} \bibinfo{person}{Yohann Paris}.} \bibinfo{year}{2021}\natexlab{}.
\newblock \showarticletitle{{ A Multi-Scale Visual Analytics Approach for Exploring Biomedical Knowledge }}. In \bibinfo{booktitle}{\emph{2021 IEEE Workshop on Visual Analytics in Healthcare (VAHC)}}. \bibinfo{publisher}{IEEE Computer Society}, \bibinfo{address}{Los Alamitos, CA, USA}, \bibinfo{pages}{30--35}.
\newblock
\href{https://doi.org/10.1109/VAHC53616.2021.00010}{doi:\nolinkurl{10.1109/VAHC53616.2021.00010}}


\bibitem[Jacomy and Plique(2014)]%
        {sigmajs}
\bibfield{author}{\bibinfo{person}{Alexis Jacomy} {and} \bibinfo{person}{Guillaume Plique}.} \bibinfo{year}{2014}\natexlab{}.
\newblock \bibinfo{title}{sigma.js: JavaScript Library for Network Visualization}.
\newblock \bibinfo{howpublished}{\url{https://www.sigmajs.org/}}.
\newblock
\newblock
\shownote{Network graph rendering library, accessed 2025-12-04}.


\bibitem[Jia et~al\mbox{.}(2025)]%
        {jia2024medikal}
\bibfield{author}{\bibinfo{person}{Mingyi Jia}, \bibinfo{person}{Junwen Duan}, \bibinfo{person}{Yan Song}, {and} \bibinfo{person}{Jianxin Wang}.} \bibinfo{year}{2025}\natexlab{}.
\newblock \showarticletitle{med{IKAL}: Integrating Knowledge Graphs as Assistants of {LLM}s for Enhanced Clinical Diagnosis on {EMR}s}. In \bibinfo{booktitle}{\emph{Proceedings of the 31st International Conference on Computational Linguistics}} (Abu Dhabi, UAE). \bibinfo{publisher}{Association for Computational Linguistics}, \bibinfo{address}{Stroudsburg, PA, USA}, \bibinfo{pages}{9278--9298}.
\newblock
\urldef\tempurl%
\url{https://aclanthology.org/2025.coling-main.624/}
\showURL{%
\tempurl}


\bibitem[Knitza et~al\mbox{.}(2021)]%
        {knitza2021accuracy}
\bibfield{author}{\bibinfo{person}{Johannes Knitza}, \bibinfo{person}{Koray Tascilar}, \bibinfo{person}{Eva Gruber}, \bibinfo{person}{Hannah Kaletta}, \bibinfo{person}{Melanie Hagen}, \bibinfo{person}{Anna-Maria Liphardt}, \bibinfo{person}{Hannah Schenker}, \bibinfo{person}{Martin Krusche}, \bibinfo{person}{Jochen Wacker}, \bibinfo{person}{Arnd Kleyer}, {et~al\mbox{.}}} \bibinfo{year}{2021}\natexlab{}.
\newblock \showarticletitle{Accuracy and usability of a diagnostic decision support system in the diagnosis of three representative rheumatic diseases: a randomized controlled trial among medical students}.
\newblock \bibinfo{journal}{\emph{Arthritis Research \& Therapy}} \bibinfo{volume}{23}, \bibinfo{number}{1}, Article \bibinfo{articleno}{233} (\bibinfo{year}{2021}), \bibinfo{numpages}{10}~pages.
\newblock
\href{https://doi.org/10.1186/s13075-021-02616-6}{doi:\nolinkurl{10.1186/s13075-021-02616-6}}


\bibitem[Kou et~al\mbox{.}(2022)]%
        {kou2022hc}
\bibfield{author}{\bibinfo{person}{Ziyi Kou}, \bibinfo{person}{Lanyu Shang}, \bibinfo{person}{Yang Zhang}, {and} \bibinfo{person}{Dong Wang}.} \bibinfo{year}{2022}\natexlab{}.
\newblock \showarticletitle{Hc-covid: A hierarchical crowdsource knowledge graph approach to explainable covid-19 misinformation detection}.
\newblock \bibinfo{journal}{\emph{Proceedings of the ACM on Human-Computer Interaction}} \bibinfo{volume}{6}, \bibinfo{number}{GROUP} (\bibinfo{year}{2022}), \bibinfo{pages}{1--25}.
\newblock
\href{https://doi.org/10.1145/3492855}{doi:\nolinkurl{10.1145/3492855}}


\bibitem[Lewis et~al\mbox{.}(2020)]%
        {lewis2020retrieval}
\bibfield{author}{\bibinfo{person}{Patrick Lewis}, \bibinfo{person}{Ethan Perez}, \bibinfo{person}{Aleksandra Piktus}, \bibinfo{person}{Fabio Petroni}, \bibinfo{person}{Vladimir Karpukhin}, \bibinfo{person}{Naman Goyal}, \bibinfo{person}{Heinrich K\"{u}ttler}, \bibinfo{person}{Mike Lewis}, \bibinfo{person}{Wen-tau Yih}, \bibinfo{person}{Tim Rockt\"{a}schel}, \bibinfo{person}{Sebastian Riedel}, {and} \bibinfo{person}{Douwe Kiela}.} \bibinfo{year}{2020}\natexlab{}.
\newblock \showarticletitle{Retrieval-augmented generation for knowledge-intensive NLP tasks}. In \bibinfo{booktitle}{\emph{Proceedings of the 34th International Conference on Neural Information Processing Systems}} (Vancouver, BC, Canada) \emph{(\bibinfo{series}{NIPS '20})}. \bibinfo{publisher}{Curran Associates Inc.}, \bibinfo{address}{Red Hook, NY, USA}, Article \bibinfo{articleno}{793}, \bibinfo{numpages}{16}~pages.
\newblock
\showISBNx{9781713829546}
\href{https://doi.org/10.5555/3495724.3496517}{doi:\nolinkurl{10.5555/3495724.3496517}}


\bibitem[Li et~al\mbox{.}(2023)]%
        {li2023knowledge}
\bibfield{author}{\bibinfo{person}{Harry Li}, \bibinfo{person}{Gabriel Appleby}, \bibinfo{person}{Camelia~Daniela Brumar}, \bibinfo{person}{Remco Chang}, {and} \bibinfo{person}{Ashley Suh}.} \bibinfo{year}{2023}\natexlab{}.
\newblock \showarticletitle{Knowledge graphs in practice: Characterizing their users, challenges, and visualization opportunities}.
\newblock \bibinfo{journal}{\emph{IEEE Transactions on Visualization and Computer Graphics}} \bibinfo{volume}{30}, \bibinfo{number}{1} (\bibinfo{year}{2023}), \bibinfo{pages}{584--594}.
\newblock
\href{https://doi.org/10.1109/TVCG.2023.3326904}{doi:\nolinkurl{10.1109/TVCG.2023.3326904}}


\bibitem[Li et~al\mbox{.}(2024)]%
        {li2024linkq}
\bibfield{author}{\bibinfo{person}{Harry Li}, \bibinfo{person}{Gabriel Appleby}, {and} \bibinfo{person}{Ashley Suh}.} \bibinfo{year}{2024}\natexlab{}.
\newblock \showarticletitle{{ LinkQ: An LLM-Assisted Visual Interface for Knowledge Graph Question-Answering }}. In \bibinfo{booktitle}{\emph{IEEE Visualization and Visual Analytics (VIS)}}. \bibinfo{publisher}{IEEE Computer Society}, \bibinfo{address}{Los Alamitos, CA, USA}, \bibinfo{pages}{116--120}.
\newblock
\href{https://doi.org/10.1109/VIS55277.2024.00031}{doi:\nolinkurl{10.1109/VIS55277.2024.00031}}


\bibitem[Li et~al\mbox{.}(2025a)]%
        {li2025accurate}
\bibfield{author}{\bibinfo{person}{Haoran Li}, \bibinfo{person}{Xusen Cheng}, {and} \bibinfo{person}{Xiaoping Zhang}.} \bibinfo{year}{2025}\natexlab{a}.
\newblock \showarticletitle{Accurate Insights, Trustworthy Interactions: Designing a Collaborative AI-Human Multi-Agent System with Knowledge Graph for Diagnosis Prediction}. In \bibinfo{booktitle}{\emph{Proceedings of the 2025 CHI Conference on Human Factors in Computing Systems}}. \bibinfo{publisher}{Association for Computing Machinery}, \bibinfo{address}{New York, NY, USA}, \bibinfo{pages}{1--15}.
\newblock
\href{https://doi.org/10.1145/3706598.3713526}{doi:\nolinkurl{10.1145/3706598.3713526}}


\bibitem[Li et~al\mbox{.}(2025b)]%
        {Li2025Automatic}
\bibfield{author}{\bibinfo{person}{Nan Li}, \bibinfo{person}{Bing Xue}, \bibinfo{person}{Lianbo Ma}, {and} \bibinfo{person}{Mengjie Zhang}.} \bibinfo{year}{2025}\natexlab{b}.
\newblock \showarticletitle{Automatic Fuzzy Architecture Design for Defect Detection via Classifier-Assisted Multiobjective Optimization Approach}.
\newblock \bibinfo{journal}{\emph{IEEE Transactions on Evolutionary Computation}} (\bibinfo{date}{jan} \bibinfo{year}{2025}), \bibinfo{pages}{1}.
\newblock
\href{https://doi.org/10.1109/TEVC.2025.3530416}{doi:\nolinkurl{10.1109/TEVC.2025.3530416}}


\bibitem[Liu et~al\mbox{.}(2022)]%
        {liu2022personal}
\bibfield{author}{\bibinfo{person}{Yinan Liu}, \bibinfo{person}{Hu Chen}, {and} \bibinfo{person}{Wei Shen}.} \bibinfo{year}{2022}\natexlab{}.
\newblock \showarticletitle{Personal attribute prediction from conversations}. In \bibinfo{booktitle}{\emph{Companion Proceedings of the Web Conference 2022}}. \bibinfo{pages}{223--227}.
\newblock


\bibitem[Liu et~al\mbox{.}(2023)]%
        {liu2023low}
\bibfield{author}{\bibinfo{person}{Yinan Liu}, \bibinfo{person}{Hu Chen}, \bibinfo{person}{Wei Shen}, {and} \bibinfo{person}{Jiaoyan Chen}.} \bibinfo{year}{2023}\natexlab{}.
\newblock \showarticletitle{Low-resource personal attribute prediction from conversations}. In \bibinfo{booktitle}{\emph{Proceedings of the AAAI Conference on Artificial Intelligence}}, Vol.~\bibinfo{volume}{37}. \bibinfo{pages}{4507--4515}.
\newblock


\bibitem[Liu et~al\mbox{.}(2021)]%
        {liu2021joint}
\bibfield{author}{\bibinfo{person}{Yinan Liu}, \bibinfo{person}{Wei Shen}, \bibinfo{person}{Yuanfei Wang}, \bibinfo{person}{Jianyong Wang}, \bibinfo{person}{Zhenglu Yang}, {and} \bibinfo{person}{Xiaojie Yuan}.} \bibinfo{year}{2021}\natexlab{}.
\newblock \showarticletitle{Joint open knowledge base canonicalization and linking}. In \bibinfo{booktitle}{\emph{Proceedings of the 2021 International Conference on Management of Data}}. \bibinfo{pages}{2253--2261}.
\newblock


\bibitem[Liu et~al\mbox{.}(2020)]%
        {liu2020named}
\bibfield{author}{\bibinfo{person}{Yinan Liu}, \bibinfo{person}{Wei Shen}, \bibinfo{person}{Zonghai Yao}, \bibinfo{person}{Jianyong Wang}, \bibinfo{person}{Zhenglu Yang}, {and} \bibinfo{person}{Xiaojie Yuan}.} \bibinfo{year}{2020}\natexlab{}.
\newblock \showarticletitle{Named entity location prediction combining twitter and web}.
\newblock \bibinfo{journal}{\emph{IEEE Transactions on Knowledge and Data Engineering}} \bibinfo{volume}{33}, \bibinfo{number}{11} (\bibinfo{year}{2020}), \bibinfo{pages}{3618--3633}.
\newblock


\bibitem[Luo et~al\mbox{.}(2025b)]%
        {luo2024graph}
\bibfield{author}{\bibinfo{person}{Linhao Luo}, \bibinfo{person}{Zicheng Zhao}, \bibinfo{person}{Gholamreza Haffari}, \bibinfo{person}{Yuan-Fang Li}, \bibinfo{person}{Chen Gong}, {and} \bibinfo{person}{Shirui Pan}.} \bibinfo{year}{2025}\natexlab{b}.
\newblock \bibinfo{title}{Graph-constrained reasoning: Faithful reasoning on knowledge graphs with large language models}.
\newblock
\href{https://doi.org/10.48550/arXiv.2410.13080}{doi:\nolinkurl{10.48550/arXiv.2410.13080}}


\bibitem[Luo et~al\mbox{.}(2025a)]%
        {luo2025etrqa}
\bibfield{author}{\bibinfo{person}{Sigang Luo}, \bibinfo{person}{Yinan Liu}, \bibinfo{person}{Dongying Lin}, \bibinfo{person}{Yingying Zhai}, \bibinfo{person}{Bin Wang}, \bibinfo{person}{Xiaochun Yang}, {and} \bibinfo{person}{Junpeng Liu}.} \bibinfo{year}{2025}\natexlab{a}.
\newblock \showarticletitle{ETRQA: A Comprehensive Benchmark for Evaluating Event Temporal Reasoning Abilities of Large Language Models}. In \bibinfo{booktitle}{\emph{Findings of the Association for Computational Linguistics: ACL 2025}}. \bibinfo{pages}{23321--23339}.
\newblock


\bibitem[Ma et~al\mbox{.}(2025b)]%
        {Ma2025Defying}
\bibfield{author}{\bibinfo{person}{Lianbo Ma}, \bibinfo{person}{Yuee Zhou}, \bibinfo{person}{Ye Ma}, \bibinfo{person}{Guo Yu}, \bibinfo{person}{Qing Li}, \bibinfo{person}{Qiang He}, {and} \bibinfo{person}{Yan Pei}.} \bibinfo{year}{2025}\natexlab{b}.
\newblock \showarticletitle{Defying Multi-Model Forgetting in One-Shot Neural Architecture Search Using Orthogonal Gradient Learning}.
\newblock \bibinfo{journal}{\emph{IEEE Trans. Comput.}} \bibinfo{volume}{74}, \bibinfo{number}{5} (\bibinfo{date}{may} \bibinfo{year}{2025}), \bibinfo{pages}{1678--1689}.
\newblock
\href{https://doi.org/10.1109/TC.2025.3540650}{doi:\nolinkurl{10.1109/TC.2025.3540650}}


\bibitem[Ma et~al\mbox{.}(2025a)]%
        {ma2024think}
\bibfield{author}{\bibinfo{person}{Shengjie Ma}, \bibinfo{person}{Chengjin Xu}, \bibinfo{person}{Xuhui Jiang}, \bibinfo{person}{Muzhi Li}, \bibinfo{person}{Huaren Qu}, {and} \bibinfo{person}{Jian Guo}.} \bibinfo{year}{2025}\natexlab{a}.
\newblock \bibinfo{title}{Think-on-Graph 2.0: Deep and Interpretable Large Language Model Reasoning with Knowledge Graph-Guided Retrieval}.
\newblock
\href{https://doi.org/10.48550/arXiv.2407.10805}{doi:\nolinkurl{10.48550/arXiv.2407.10805}}


\bibitem[Marchesin and Silvello(2024)]%
        {marchesin_silvello-vldb2024}
\bibfield{author}{\bibinfo{person}{S. Marchesin} {and} \bibinfo{person}{G. Silvello}.} \bibinfo{year}{2024}\natexlab{}.
\newblock \showarticletitle{Efficient and Reliable Estimation of Knowledge Graph Accuracy}.
\newblock \bibinfo{journal}{\emph{Proc. {VLDB} Endow.}} \bibinfo{volume}{17}, \bibinfo{number}{9} (\bibinfo{year}{2024}), \bibinfo{pages}{2392--2404}.
\newblock
\href{https://doi.org/10.14778/3665844.3665865}{doi:\nolinkurl{10.14778/3665844.3665865}}


\bibitem[Meng et~al\mbox{.}(2025)]%
        {meng2025deconstructing}
\bibfield{author}{\bibinfo{person}{Han Meng}, \bibinfo{person}{Renwen Zhang}, \bibinfo{person}{Ganyi Wang}, \bibinfo{person}{Yitian Yang}, \bibinfo{person}{Peinuan Qin}, \bibinfo{person}{Jungup Lee}, {and} \bibinfo{person}{Yi-Chieh Lee}.} \bibinfo{year}{2025}\natexlab{}.
\newblock \showarticletitle{Deconstructing Depression Stigma: Integrating AI-driven Data Collection and Analysis with Causal Knowledge Graphs}. In \bibinfo{booktitle}{\emph{CHI 2025 - Proceedings of the 2025 CHI Conference on Human Factors in Computing Systems}} \emph{(\bibinfo{series}{Conference on Human Factors in Computing Systems - Proceedings})}. \bibinfo{publisher}{Association for Computing Machinery}, \bibinfo{address}{Yokohama, Japan}.
\newblock
\showISBNx{979-8-4007-1394-1}
\href{https://doi.org/10.1145/3706598.3714255}{doi:\nolinkurl{10.1145/3706598.3714255}}


\bibitem[Morey et~al\mbox{.}(2025a)]%
        {morey2025empirically}
\bibfield{author}{\bibinfo{person}{Dane~A Morey}, \bibinfo{person}{Michael~F Rayo}, {and} \bibinfo{person}{David~D Woods}.} \bibinfo{year}{2025}\natexlab{a}.
\newblock \showarticletitle{Empirically derived evaluation requirements for responsible deployments of AI in safety-critical settings}.
\newblock \bibinfo{journal}{\emph{npj Digital Medicine}} \bibinfo{volume}{8}, \bibinfo{number}{1} (\bibinfo{year}{2025}), \bibinfo{pages}{374}.
\newblock
\href{https://doi.org/10.1038/s41746-025-01784-y}{doi:\nolinkurl{10.1038/s41746-025-01784-y}}


\bibitem[Morey et~al\mbox{.}(2025b)]%
        {Morey2025Evaluation}
\bibfield{author}{\bibinfo{person}{Dane~A. Morey}, \bibinfo{person}{Michael~F. Rayo}, {and} \bibinfo{person}{David~D. Woods}.} \bibinfo{year}{2025}\natexlab{b}.
\newblock \showarticletitle{Empirically Derived Evaluation Requirements for Responsible Deployments of {AI} in Safety{-}Critical Settings}.
\newblock \bibinfo{journal}{\emph{npj Digital Medicine}} \bibinfo{volume}{8}, \bibinfo{number}{1} (\bibinfo{year}{2025}), \bibinfo{pages}{374}.
\newblock
\href{https://doi.org/10.1038/s41746-025-01784-y}{doi:\nolinkurl{10.1038/s41746-025-01784-y}}


\bibitem[Musen et~al\mbox{.}(2021)]%
        {musen2021clinical}
\bibfield{author}{\bibinfo{person}{Mark~A Musen}, \bibinfo{person}{Blackford Middleton}, {and} \bibinfo{person}{Robert~A Greenes}.} \bibinfo{year}{2021}\natexlab{}.
\newblock \showarticletitle{Clinical decision-support systems}.
\newblock In \bibinfo{booktitle}{\emph{Biomedical informatics: computer applications in health care and biomedicine}}. \bibinfo{publisher}{Springer}, \bibinfo{address}{Cham}, \bibinfo{pages}{795--840}.
\newblock
\href{https://doi.org/10.1007/978-3-030-58721-5_24}{doi:\nolinkurl{10.1007/978-3-030-58721-5_24}}


\bibitem[Nelson et~al\mbox{.}(2022)]%
        {Nelson2022SPOKEsig}
\bibfield{author}{\bibinfo{person}{Charlotte~A. Nelson}, \bibinfo{person}{Riley Bove}, \bibinfo{person}{Atul~J. Butte}, {and} \bibinfo{person}{Sergio~E. Baranzini}.} \bibinfo{year}{2022}\natexlab{}.
\newblock \showarticletitle{Embedding electronic health records onto a knowledge network recognizes prodromal features of multiple sclerosis and predicts diagnosis}.
\newblock \bibinfo{journal}{\emph{Journal of the American Medical Informatics Association}} \bibinfo{volume}{29}, \bibinfo{number}{3} (\bibinfo{year}{2022}), \bibinfo{pages}{424--434}.
\newblock
\href{https://doi.org/10.1093/jamia/ocab270}{doi:\nolinkurl{10.1093/jamia/ocab270}}


\bibitem[{Neo4j Inc.}(2010)]%
        {neo4j}
\bibfield{author}{\bibinfo{person}{{Neo4j Inc.}}} \bibinfo{year}{2010}\natexlab{}.
\newblock \bibinfo{title}{Neo4j Graph Database}.
\newblock \bibinfo{howpublished}{\url{https://neo4j.com/}}.
\newblock
\newblock
\shownote{Neo4j graph database platform, accessed 2025-12-04}.


\bibitem[Newman‑Toker et~al\mbox{.}(2024)]%
        {NewmanToker2024}
\bibfield{author}{\bibinfo{person}{David~E. Newman‑Toker}, \bibinfo{person}{Najlla Nassery}, \bibinfo{person}{Adam~C. Schaffer}, \bibinfo{person}{Chihwen~Winnie Yu‑Moe}, \bibinfo{person}{Gwendolyn~D. Clemens}, \bibinfo{person}{Zheyu Wang}, \bibinfo{person}{Yuxin Zhu}, \bibinfo{person}{Ali~S. Saber~Tehrani}, \bibinfo{person}{Mehdi Fanai}, \bibinfo{person}{Ahmed Hassoon}, {and} \bibinfo{person}{Dana Siegal}.} \bibinfo{year}{2024}\natexlab{}.
\newblock \showarticletitle{Burden of serious harms from diagnostic error in the USA}.
\newblock \bibinfo{journal}{\emph{BMJ Quality \& Safety}} \bibinfo{volume}{33}, \bibinfo{number}{2} (\bibinfo{year}{2024}), \bibinfo{pages}{109--120}.
\newblock
\href{https://doi.org/10.1136/bmjqs-2021-014130}{doi:\nolinkurl{10.1136/bmjqs-2021-014130}}


\bibitem[Nguyen et~al\mbox{.}(2024)]%
        {Nguyen2024ChatbotFHx}
\bibfield{author}{\bibinfo{person}{Michelle~Hoang Nguyen}, \bibinfo{person}{Jo{\~a}o Sedoc}, {and} \bibinfo{person}{Casey~Overby Taylor}.} \bibinfo{year}{2024}\natexlab{}.
\newblock \showarticletitle{Usability, Engagement, and Report Usefulness of Chatbot{-}Based Family Health History Data Collection: Mixed Methods Analysis}.
\newblock \bibinfo{journal}{\emph{Journal of Medical Internet Research}}  \bibinfo{volume}{26} (\bibinfo{year}{2024}), \bibinfo{pages}{e55164}.
\newblock
\href{https://doi.org/10.2196/55164}{doi:\nolinkurl{10.2196/55164}}


\bibitem[Oami et~al\mbox{.}(2024)]%
        {Oami2024JAMANetwOpen}
\bibfield{author}{\bibinfo{person}{Takehiko Oami}, \bibinfo{person}{Yohei Okada}, {and} \bibinfo{person}{Taka-Aki Nakada}.} \bibinfo{year}{2024}\natexlab{}.
\newblock \showarticletitle{Performance of a Large Language Model in Screening Citations}.
\newblock \bibinfo{journal}{\emph{JAMA Network Open}} \bibinfo{volume}{7}, \bibinfo{number}{7} (\bibinfo{date}{Jul} \bibinfo{year}{2024}), \bibinfo{pages}{e2420496}.
\newblock
\href{https://doi.org/10.1001/jamanetworkopen.2024.20496}{doi:\nolinkurl{10.1001/jamanetworkopen.2024.20496}}


\bibitem[Oh et~al\mbox{.}(2022)]%
        {oh2022_pgd_asymmetry}
\bibfield{author}{\bibinfo{person}{Chi~Young Oh}, \bibinfo{person}{Yuhan Luo}, \bibinfo{person}{Beth {St. Jean}}, {and} \bibinfo{person}{Eun~Kyoung Choe}.} \bibinfo{year}{2022}\natexlab{}.
\newblock \showarticletitle{Patients Waiting for Cues: Information Asymmetries and Challenges in Sharing Patient-Generated Data in the Clinic}.
\newblock \bibinfo{journal}{\emph{Proceedings of the ACM on Human-Computer Interaction}} \bibinfo{volume}{6}, \bibinfo{number}{CSCW1} (\bibinfo{year}{2022}), \bibinfo{pages}{107:1--107:23}.
\newblock
\href{https://doi.org/10.1145/3512954}{doi:\nolinkurl{10.1145/3512954}}


\bibitem[{OpenAI}(2025)]%
        {gpt41}
\bibfield{author}{\bibinfo{person}{{OpenAI}}.} \bibinfo{year}{2025}\natexlab{}.
\newblock \bibinfo{title}{GPT-4.1 in the OpenAI API}.
\newblock \bibinfo{howpublished}{\url{https://openai.com/index/gpt-4-1/}}.
\newblock
\newblock
\shownote{Model overview and release announcement, accessed 2025-12-04}.


\bibitem[Ouanes and Farhah(2024)]%
        {ouanes2024effectiveness}
\bibfield{author}{\bibinfo{person}{Khaled Ouanes} {and} \bibinfo{person}{Nesren Farhah}.} \bibinfo{year}{2024}\natexlab{}.
\newblock \showarticletitle{Effectiveness of artificial intelligence (AI) in clinical decision support systems and care delivery}.
\newblock \bibinfo{journal}{\emph{Journal of medical systems}} \bibinfo{volume}{48}, \bibinfo{number}{1} (\bibinfo{year}{2024}), \bibinfo{pages}{74}.
\newblock
\href{https://doi.org/10.1007/s10916-024-02098-4}{doi:\nolinkurl{10.1007/s10916-024-02098-4}}


\bibitem[Plique(2016)]%
        {graphology}
\bibfield{author}{\bibinfo{person}{Guillaume Plique}.} \bibinfo{year}{2016}\natexlab{}.
\newblock \bibinfo{title}{Graphology: A JavaScript Graph Library}.
\newblock \bibinfo{howpublished}{\url{https://graphology.github.io/}}.
\newblock
\newblock
\shownote{medialab Sciences Po graph library, accessed 2025-12-04}.


\bibitem[Riley et~al\mbox{.}(2017)]%
        {riley2017care}
\bibfield{author}{\bibinfo{person}{David~S Riley}, \bibinfo{person}{Melissa~S Barber}, \bibinfo{person}{Gunver~S Kienle}, \bibinfo{person}{Jeffrey~K Aronson}, \bibinfo{person}{Tido von Schoen-Angerer}, \bibinfo{person}{Peter Tugwell}, \bibinfo{person}{Helmut Kiene}, \bibinfo{person}{Mark Helfand}, \bibinfo{person}{Douglas~G Altman}, \bibinfo{person}{Harold Sox}, {et~al\mbox{.}}} \bibinfo{year}{2017}\natexlab{}.
\newblock \showarticletitle{CARE guidelines for case reports: explanation and elaboration document}.
\newblock \bibinfo{journal}{\emph{Journal of clinical epidemiology}}  \bibinfo{volume}{89} (\bibinfo{year}{2017}), \bibinfo{pages}{218--235}.
\newblock
\href{https://doi.org/10.1016/j.jclinepi.2017.04.026}{doi:\nolinkurl{10.1016/j.jclinepi.2017.04.026}}


\bibitem[Ronacher and Team(2025)]%
        {flask}
\bibfield{author}{\bibinfo{person}{Armin Ronacher} {and} \bibinfo{person}{The~Pallets Team}.} \bibinfo{year}{2025}\natexlab{}.
\newblock \bibinfo{title}{Flask Documentation}.
\newblock \bibinfo{howpublished}{\url{https://flask.palletsprojects.com/}}.
\newblock
\newblock
\shownote{Flask web framework documentation, accessed 2025-12-04}.


\bibitem[Rotmensch et~al\mbox{.}(2017)]%
        {rotmensch2017learning}
\bibfield{author}{\bibinfo{person}{Maya Rotmensch}, \bibinfo{person}{Yoni Halpern}, \bibinfo{person}{Abdulhakim Tlimat}, \bibinfo{person}{Steven Horng}, {and} \bibinfo{person}{David Sontag}.} \bibinfo{year}{2017}\natexlab{}.
\newblock \showarticletitle{Learning a health knowledge graph from electronic medical records}.
\newblock \bibinfo{journal}{\emph{Scientific reports}} \bibinfo{volume}{7}, \bibinfo{number}{1}, Article \bibinfo{articleno}{5994} (\bibinfo{year}{2017}), \bibinfo{numpages}{11}~pages.
\newblock
\href{https://doi.org/10.1038/s41598-017-05778-z}{doi:\nolinkurl{10.1038/s41598-017-05778-z}}


\bibitem[Roy and Pan(2021)]%
        {roy2021incorporating}
\bibfield{author}{\bibinfo{person}{Arpita Roy} {and} \bibinfo{person}{Shimei Pan}.} \bibinfo{year}{2021}\natexlab{}.
\newblock \showarticletitle{Incorporating medical knowledge in {BERT} for clinical relation extraction}. In \bibinfo{booktitle}{\emph{Proceedings of the 2021 Conference on Empirical Methods in Natural Language Processing}}. \bibinfo{publisher}{Association for Computational Linguistics}, \bibinfo{address}{Online and Punta Cana, Dominican Republic}, \bibinfo{pages}{5357--5366}.
\newblock
\href{https://doi.org/10.18653/v1/2021.emnlp-main.435}{doi:\nolinkurl{10.18653/v1/2021.emnlp-main.435}}


\bibitem[Sauro and Lewis(2016)]%
        {SauroLewis2016QuantifyingUX}
\bibfield{author}{\bibinfo{person}{Jeff Sauro} {and} \bibinfo{person}{James~R. Lewis}.} \bibinfo{year}{2016}\natexlab{}.
\newblock \bibinfo{booktitle}{\emph{Quantifying the User Experience: Practical Statistics for User Research}}.
\newblock \bibinfo{publisher}{Morgan Kaufmann}, \bibinfo{address}{Waltham, MA}.
\newblock
\showISBNx{9780128023082}


\bibitem[Shang et~al\mbox{.}(2024)]%
        {shang2024electronic}
\bibfield{author}{\bibinfo{person}{Yong Shang}, \bibinfo{person}{Yu Tian}, \bibinfo{person}{Kewei Lyu}, \bibinfo{person}{Tianshu Zhou}, \bibinfo{person}{Ping Zhang}, \bibinfo{person}{Jianghua Chen}, {and} \bibinfo{person}{Jingsong Li}.} \bibinfo{year}{2024}\natexlab{}.
\newblock \showarticletitle{Electronic health record--oriented knowledge graph system for collaborative clinical decision support using multicenter fragmented medical data: design and application study}.
\newblock \bibinfo{journal}{\emph{Journal of Medical Internet Research}}  \bibinfo{volume}{26}, Article \bibinfo{articleno}{e54263} (\bibinfo{year}{2024}), \bibinfo{numpages}{13}~pages.
\newblock
\href{https://doi.org/10.2196/54263}{doi:\nolinkurl{10.2196/54263}}


\bibitem[Shen et~al\mbox{.}(2018)]%
        {shen2018predicting}
\bibfield{author}{\bibinfo{person}{Wei Shen}, \bibinfo{person}{Yinan Liu}, {and} \bibinfo{person}{Jianyong Wang}.} \bibinfo{year}{2018}\natexlab{}.
\newblock \showarticletitle{Predicting named entity location using Twitter}. In \bibinfo{booktitle}{\emph{2018 IEEE 34th International Conference on Data Engineering (ICDE)}}. IEEE, \bibinfo{pages}{161--172}.
\newblock


\bibitem[Shneiderman(2020)]%
        {Shneiderman2020HCAI}
\bibfield{author}{\bibinfo{person}{Ben Shneiderman}.} \bibinfo{year}{2020}\natexlab{}.
\newblock \showarticletitle{Human{-}Centered Artificial Intelligence: Reliable, Safe \& Trustworthy}.
\newblock \bibinfo{journal}{\emph{International Journal of Human{-}Computer Interaction}} \bibinfo{volume}{36}, \bibinfo{number}{6} (\bibinfo{year}{2020}), \bibinfo{pages}{495--504}.
\newblock
\href{https://doi.org/10.1080/10447318.2020.1741118}{doi:\nolinkurl{10.1080/10447318.2020.1741118}}


\bibitem[sidebase(2025)]%
        {nuxtauth}
\bibfield{author}{\bibinfo{person}{sidebase}.} \bibinfo{year}{2025}\natexlab{}.
\newblock \bibinfo{title}{@sidebase/nuxt-auth}.
\newblock \bibinfo{howpublished}{\url{https://github.com/sidebase/nuxt-auth}}.
\newblock
\newblock
\shownote{Authentication module for Nuxt 3, accessed 2025-12-04}.


\bibitem[Sokol et~al\mbox{.}(2025a)]%
        {sokol2025artificial}
\bibfield{author}{\bibinfo{person}{Kacper Sokol}, \bibinfo{person}{James Fackler}, {and} \bibinfo{person}{Julia~E Vogt}.} \bibinfo{year}{2025}\natexlab{a}.
\newblock \showarticletitle{Artificial intelligence should genuinely support clinical reasoning and decision making to bridge the translational gap}.
\newblock \bibinfo{journal}{\emph{npj Digital Medicine}} \bibinfo{volume}{8}, \bibinfo{number}{1} (\bibinfo{year}{2025}), \bibinfo{pages}{345}.
\newblock
\href{https://doi.org/10.1038/s41746-025-01725-9}{doi:\nolinkurl{10.1038/s41746-025-01725-9}}


\bibitem[Sokol et~al\mbox{.}(2025b)]%
        {Sokol2025ClinicalReasoning}
\bibfield{author}{\bibinfo{person}{Kacper Sokol}, \bibinfo{person}{James Fackler}, {and} \bibinfo{person}{Julia~E. Vogt}.} \bibinfo{year}{2025}\natexlab{b}.
\newblock \showarticletitle{Artificial Intelligence Should Genuinely Support Clinical Reasoning and Decision Making to Bridge the Translational Gap}.
\newblock \bibinfo{journal}{\emph{npj Digital Medicine}} \bibinfo{volume}{8}, \bibinfo{number}{1} (\bibinfo{year}{2025}), \bibinfo{pages}{345}.
\newblock
\href{https://doi.org/10.1038/s41746-025-01725-9}{doi:\nolinkurl{10.1038/s41746-025-01725-9}}


\bibitem[Sreekar et~al\mbox{.}(2024)]%
        {sreekar2024axcel}
\bibfield{author}{\bibinfo{person}{P~Aditya Sreekar}, \bibinfo{person}{Sahil Verma}, \bibinfo{person}{Suransh Chopra}, \bibinfo{person}{Sarik Ghazarian}, \bibinfo{person}{Abhishek Persad}, {and} \bibinfo{person}{Narayanan Sadagopan}.} \bibinfo{year}{2024}\natexlab{}.
\newblock \showarticletitle{AXCEL: Automated eXplainable consistency evaluation using LLMs}. In \bibinfo{booktitle}{\emph{Findings of the Association for Computational Linguistics: EMNLP 2024}} (Bangkok, Thailand). \bibinfo{publisher}{Association for Computational Linguistics}, \bibinfo{address}{Stroudsburg, PA, USA}, \bibinfo{pages}{14943--14957}.
\newblock
\href{https://doi.org/10.18653/v1/2024.findings-emnlp.878}{doi:\nolinkurl{10.18653/v1/2024.findings-emnlp.878}}


\bibitem[Sun et~al\mbox{.}(2020)]%
        {sun2020dfseer}
\bibfield{author}{\bibinfo{person}{Dong Sun}, \bibinfo{person}{Zezheng Feng}, \bibinfo{person}{Yuanzhe Chen}, \bibinfo{person}{Yong Wang}, \bibinfo{person}{Jia Zeng}, \bibinfo{person}{Mingxuan Yuan}, \bibinfo{person}{Ting-Chuen Pong}, {and} \bibinfo{person}{Huamin Qu}.} \bibinfo{year}{2020}\natexlab{}.
\newblock \showarticletitle{DfSeer: A Visual Analytics Approach to Facilitate Model Selection for Demand Forecasting}. In \bibinfo{booktitle}{\emph{Proceedings of the 2020 {CHI} Conference on Human Factors in Computing Systems}}. \bibinfo{pages}{1--13}.
\newblock


\bibitem[Sun et~al\mbox{.}(2024)]%
        {sun2023think}
\bibfield{author}{\bibinfo{person}{Jiashuo Sun}, \bibinfo{person}{Chengjin Xu}, \bibinfo{person}{Lumingyuan Tang}, \bibinfo{person}{Saizhuo Wang}, \bibinfo{person}{Chen Lin}, \bibinfo{person}{Yeyun Gong}, \bibinfo{person}{Lionel~M. Ni}, \bibinfo{person}{Heung-Yeung Shum}, {and} \bibinfo{person}{Jian Guo}.} \bibinfo{year}{2024}\natexlab{}.
\newblock \bibinfo{title}{Think-on-Graph: Deep and Responsible Reasoning of Large Language Models on Knowledge Graph}.
\newblock
\href{https://doi.org/10.48550/arXiv.2307.07697}{doi:\nolinkurl{10.48550/arXiv.2307.07697}}


\bibitem[Sutton et~al\mbox{.}(2020)]%
        {sutton2020overview}
\bibfield{author}{\bibinfo{person}{Reed~T Sutton}, \bibinfo{person}{David Pincock}, \bibinfo{person}{Daniel~C Baumgart}, \bibinfo{person}{Daniel~C Sadowski}, \bibinfo{person}{Richard~N Fedorak}, {and} \bibinfo{person}{Karen~I Kroeker}.} \bibinfo{year}{2020}\natexlab{}.
\newblock \showarticletitle{An overview of clinical decision support systems: benefits, risks, and strategies for success}.
\newblock \bibinfo{journal}{\emph{NPJ digital medicine}} \bibinfo{volume}{3}, \bibinfo{number}{1}, Article \bibinfo{articleno}{17} (\bibinfo{year}{2020}), \bibinfo{numpages}{13}~pages.
\newblock
\href{https://doi.org/10.1038/s41746-020-0221-y}{doi:\nolinkurl{10.1038/s41746-020-0221-y}}


\bibitem[Sweller(2023)]%
        {Sweller2023CLTExpansion}
\bibfield{author}{\bibinfo{person}{John Sweller}.} \bibinfo{year}{2023}\natexlab{}.
\newblock \showarticletitle{The Development of Cognitive Load Theory: Replication Crises and Incorporation of Other Theories Can Lead to Theory Expansion}.
\newblock \bibinfo{journal}{\emph{Educational Psychology Review}} \bibinfo{volume}{35}, \bibinfo{number}{4}, Article \bibinfo{articleno}{95} (\bibinfo{year}{2023}), \bibinfo{numpages}{20}~pages.
\newblock
\href{https://doi.org/10.1007/s10648-023-09817-2}{doi:\nolinkurl{10.1007/s10648-023-09817-2}}


\bibitem[Symington(2025)]%
        {Symington2025}
\bibfield{author}{\bibinfo{person}{Banu~E. Symington}.} \bibinfo{year}{2025}\natexlab{}.
\newblock \showarticletitle{Uncovering the hidden drivers of rural health care disparities}.
\newblock \bibinfo{journal}{\emph{CA: A Cancer Journal for Clinicians}} \bibinfo{volume}{75}, \bibinfo{number}{4} (\bibinfo{date}{Jul} \bibinfo{year}{2025}), \bibinfo{pages}{280--281}.
\newblock
\href{https://doi.org/10.3322/caac.70009}{doi:\nolinkurl{10.3322/caac.70009}}


\bibitem[Tang et~al\mbox{.}(2024)]%
        {Tang2024NatAging}
\bibfield{author}{\bibinfo{person}{Alice~S. Tang}, \bibinfo{person}{Katherine~P. Rankin}, \bibinfo{person}{Gabriel Cerono}, \bibinfo{person}{Silvia Miramontes}, \bibinfo{person}{Hunter Mills}, \bibinfo{person}{Jacquelyn Roger}, \bibinfo{person}{Billy Zeng}, \bibinfo{person}{Charlotte Nelson}, \bibinfo{person}{Karthik Soman}, \bibinfo{person}{Sarah Woldemariam}, \bibinfo{person}{Yaqiao Li}, \bibinfo{person}{Albert Lee}, \bibinfo{person}{Riley Bove}, \bibinfo{person}{Maria Glymour}, \bibinfo{person}{Nima Aghaeepour}, \bibinfo{person}{Tomiko~T. Oskotsky}, \bibinfo{person}{Zachary Miller}, \bibinfo{person}{Isabel~E. Allen}, \bibinfo{person}{Stephan~J. Sanders}, \bibinfo{person}{Sergio~E. Baranzini}, {and} \bibinfo{person}{Marina Sirota}.} \bibinfo{year}{2024}\natexlab{}.
\newblock \showarticletitle{Leveraging electronic health records and knowledge networks for {Alzheimer's} disease prediction and sex-specific biological insights}.
\newblock \bibinfo{journal}{\emph{Nature Aging}} \bibinfo{volume}{4}, \bibinfo{number}{3} (\bibinfo{date}{Mar} \bibinfo{year}{2024}), \bibinfo{pages}{379--395}.
\newblock
\href{https://doi.org/10.1038/s43587-024-00573-8}{doi:\nolinkurl{10.1038/s43587-024-00573-8}}


\bibitem[Tu et~al\mbox{.}(2025)]%
        {tu2025towards}
\bibfield{author}{\bibinfo{person}{Tao Tu}, \bibinfo{person}{Mike Schaekermann}, \bibinfo{person}{Anil Palepu}, \bibinfo{person}{Khaled Saab}, \bibinfo{person}{Jan Freyberg}, \bibinfo{person}{Ryutaro Tanno}, \bibinfo{person}{Amy Wang}, \bibinfo{person}{Brenna Li}, \bibinfo{person}{Mohamed Amin}, \bibinfo{person}{Yong Cheng}, \bibinfo{person}{Elahe Vedadi}, \bibinfo{person}{Nenad Tomasev}, \bibinfo{person}{Shekoofeh Azizi}, \bibinfo{person}{Karan Singhal}, \bibinfo{person}{Le Hou}, \bibinfo{person}{Albert Webson}, \bibinfo{person}{Kavita Kulkarni}, \bibinfo{person}{S.~Sara Mahdavi}, \bibinfo{person}{Christopher Semturs}, \bibinfo{person}{Juraj Gottweis}, \bibinfo{person}{Joelle Barral}, \bibinfo{person}{Katherine Chou}, \bibinfo{person}{Greg~S. Corrado}, \bibinfo{person}{Yossi Matias}, \bibinfo{person}{Alan Karthikesalingam}, {and} \bibinfo{person}{Vivek Natarajan}.} \bibinfo{year}{2025}\natexlab{}.
\newblock \showarticletitle{Towards conversational diagnostic artificial intelligence}.
\newblock \bibinfo{journal}{\emph{Nature}} \bibinfo{volume}{642}, \bibinfo{number}{8067} (\bibinfo{date}{Apr} \bibinfo{year}{2025}), \bibinfo{pages}{442--450}.
\newblock
\href{https://doi.org/10.1038/s41586-025-08866-7}{doi:\nolinkurl{10.1038/s41586-025-08866-7}}


\bibitem[Wallace et~al\mbox{.}(2022)]%
        {wallace2022diagnostic}
\bibfield{author}{\bibinfo{person}{William Wallace}, \bibinfo{person}{Calvin Chan}, \bibinfo{person}{Swathikan Chidambaram}, \bibinfo{person}{Lydia Hanna}, \bibinfo{person}{Fahad~Mujtaba Iqbal}, \bibinfo{person}{Amish Acharya}, \bibinfo{person}{Pasha Normahani}, \bibinfo{person}{Hutan Ashrafian}, \bibinfo{person}{Sheraz~R Markar}, \bibinfo{person}{Viknesh Sounderajah}, {et~al\mbox{.}}} \bibinfo{year}{2022}\natexlab{}.
\newblock \showarticletitle{The diagnostic and triage accuracy of digital and online symptom checker tools: a systematic review}.
\newblock \bibinfo{journal}{\emph{NPJ digital medicine}} \bibinfo{volume}{5}, \bibinfo{number}{1}, Article \bibinfo{articleno}{118} (\bibinfo{year}{2022}), \bibinfo{numpages}{15}~pages.
\newblock
\href{https://doi.org/10.1038/s41746-022-00667-w}{doi:\nolinkurl{10.1038/s41746-022-00667-w}}


\bibitem[Wen et~al\mbox{.}(2024)]%
        {wen2023mindmap}
\bibfield{author}{\bibinfo{person}{Yilin Wen}, \bibinfo{person}{Zifeng Wang}, {and} \bibinfo{person}{Jimeng Sun}.} \bibinfo{year}{2024}\natexlab{}.
\newblock \showarticletitle{Mindmap: Knowledge graph prompting sparks graph of thoughts in large language models}. In \bibinfo{booktitle}{\emph{Proceedings of the 62nd Annual Meeting of the Association for Computational Linguistics (ACL 2024)}} (Bangkok, Thailand). \bibinfo{publisher}{Association for Computational Linguistics}, \bibinfo{address}{Bangkok, Thailand}, \bibinfo{pages}{10370--10388}.
\newblock
\href{https://doi.org/10.18653/v1/2024.acl-long.558}{doi:\nolinkurl{10.18653/v1/2024.acl-long.558}}


\bibitem[Wilson et~al\mbox{.}(2018)]%
        {Wilson2018PALM}
\bibfield{author}{\bibinfo{person}{Michael~L. Wilson}, \bibinfo{person}{Kenneth~A. Fleming}, \bibinfo{person}{Modupe~A. Kuti}, \bibinfo{person}{Lai~Meng Looi}, \bibinfo{person}{Nestor Lago}, {and} \bibinfo{person}{Kun Ru}.} \bibinfo{year}{2018}\natexlab{}.
\newblock \showarticletitle{Access to pathology and laboratory medicine services: a crucial gap}.
\newblock \bibinfo{journal}{\emph{The Lancet}} \bibinfo{volume}{391}, \bibinfo{number}{10133} (\bibinfo{year}{2018}), \bibinfo{pages}{1927--1938}.
\newblock
\href{https://doi.org/10.1016/S0140-6736(18)30458-6}{doi:\nolinkurl{10.1016/S0140-6736(18)30458-6}}


\bibitem[{World Health Organization}(2021)]%
        {WHO2021RuralGuideline}
\bibfield{author}{\bibinfo{person}{{World Health Organization}}.} \bibinfo{year}{2021}\natexlab{}.
\newblock \bibinfo{booktitle}{\emph{WHO guideline on health workforce development, attraction, recruitment and retention in rural and remote areas}}.
\newblock \bibinfo{publisher}{World Health Organization}, \bibinfo{address}{Geneva}.
\newblock
\showISBNx{978-92-4-002422-9}
\urldef\tempurl%
\url{https://www.ncbi.nlm.nih.gov/books/NBK570763/}
\showURL{%
\tempurl}


\bibitem[{World Health Organization}(2024)]%
        {WHO2024EB156}
\bibfield{author}{\bibinfo{person}{{World Health Organization}}.} \bibinfo{year}{2024}\natexlab{}.
\newblock \bibinfo{booktitle}{\emph{Health and care workforce – Global strategy on human resources for health: workforce 2030}}.
\newblock \bibinfo{type}{Report to the Executive Board} EB156/15. \bibinfo{institution}{World Health Organization}, \bibinfo{address}{Geneva}.
\newblock
\urldef\tempurl%
\url{https://apps.who.int/gb/ebwha/pdf_files/EB156/B156_15-en.pdf}
\showURL{%
\tempurl}


\bibitem[{World Health Organization}(2025)]%
        {WHO2025PhysicianDensity}
\bibfield{author}{\bibinfo{person}{{World Health Organization}}.} \bibinfo{year}{2025}\natexlab{}.
\newblock \bibinfo{title}{Density of physicians (per 10,000 population) [Indicator]}.
\newblock \bibinfo{howpublished}{World Health Organization Global Health Observatory, dataset}.
\newblock
\urldef\tempurl%
\url{https://data.who.int/indicators/i/CCCEBB2/217795A}
\showURL{%
Retrieved 2025-09-02 from \tempurl}


\bibitem[Wu et~al\mbox{.}(2025a)]%
        {wu2025medreason}
\bibfield{author}{\bibinfo{person}{Juncheng Wu}, \bibinfo{person}{Wenlong Deng}, \bibinfo{person}{Xingxuan Li}, \bibinfo{person}{Sheng Liu}, \bibinfo{person}{Taomian Mi}, \bibinfo{person}{Yifan Peng}, \bibinfo{person}{Ziyang Xu}, \bibinfo{person}{Yi Liu}, \bibinfo{person}{Hyunjin Cho}, \bibinfo{person}{Chang-In Choi}, \bibinfo{person}{Yihan Cao}, \bibinfo{person}{Hui Ren}, \bibinfo{person}{Xiang Li}, \bibinfo{person}{Xiaoxiao Li}, {and} \bibinfo{person}{Yuyin Zhou}.} \bibinfo{year}{2025}\natexlab{a}.
\newblock \bibinfo{title}{Medreason: Eliciting factual medical reasoning steps in llms via knowledge graphs}.
\newblock
\showeprint[arxiv]{2504.00993}~[cs.AI]
\href{https://doi.org/10.48550/arXiv.2504.00993}{doi:\nolinkurl{10.48550/arXiv.2504.00993}}


\bibitem[Wu et~al\mbox{.}(2025b)]%
        {wu2025medical}
\bibfield{author}{\bibinfo{person}{Junde Wu}, \bibinfo{person}{Jiayuan Zhu}, \bibinfo{person}{Yunli Qi}, \bibinfo{person}{Jingkun Chen}, \bibinfo{person}{Min Xu}, \bibinfo{person}{Filippo Menolascina}, \bibinfo{person}{Yueming Jin}, {and} \bibinfo{person}{Vicente Grau}.} \bibinfo{year}{2025}\natexlab{b}.
\newblock \showarticletitle{Medical Graph RAG: Evidence-based Medical Large Language Model via Graph Retrieval-Augmented Generation}. In \bibinfo{booktitle}{\emph{Proceedings of the 63rd Annual Meeting of the Association for Computational Linguistics (Volume 1: Long Papers)}}. \bibinfo{publisher}{Association for Computational Linguistics}, \bibinfo{address}{Vienna, Austria}, \bibinfo{pages}{28443--28467}.
\newblock
\showISBNx{979-8-89176-251-0}
\href{https://doi.org/10.18653/v1/2025.acl-long.1381}{doi:\nolinkurl{10.18653/v1/2025.acl-long.1381}}


\bibitem[Xin and Chen(2024)]%
        {xin2024kartgps}
\bibfield{author}{\bibinfo{person}{Hao Xin} {and} \bibinfo{person}{Lei Chen}.} \bibinfo{year}{2024}\natexlab{}.
\newblock \showarticletitle{KartGPS: Knowledge Base Update with Temporal Graph Pattern-based Semantic Rules}. In \bibinfo{booktitle}{\emph{2024 IEEE 40th International Conference on Data Engineering (ICDE)}}. \bibinfo{publisher}{IEEE}, \bibinfo{address}{Los Alamitos, CA, USA}, \bibinfo{pages}{5075--5087}.
\newblock
\href{https://doi.org/10.1109/ICDE60146.2024.00105}{doi:\nolinkurl{10.1109/ICDE60146.2024.00105}}


\bibitem[Xu et~al\mbox{.}(2025a)]%
        {xu2025interactive}
\bibfield{author}{\bibinfo{person}{Jiawei Xu}, \bibinfo{person}{Juichien Chen}, \bibinfo{person}{Yilin Ye}, \bibinfo{person}{Zhandos Sembay}, \bibinfo{person}{Swathi Thaker}, \bibinfo{person}{Pamela Payne-Foster}, \bibinfo{person}{Jake~Yue Chen}, {and} \bibinfo{person}{Ying Ding}.} \bibinfo{year}{2025}\natexlab{a}.
\newblock \showarticletitle{Interactive Graph Visualization and Teaming Recommendation in an Interdisciplinary Project's Talent Knowledge Graph}.
\newblock \bibinfo{journal}{\emph{Proceedings of the Association for Information Science and Technology}} \bibinfo{volume}{62}, \bibinfo{number}{1} (\bibinfo{year}{2025}), \bibinfo{pages}{1138--1143}.
\newblock
\showISSN{2373-9231}
\href{https://doi.org/10.1002/pra2.1359}{doi:\nolinkurl{10.1002/pra2.1359}}


\bibitem[Xu et~al\mbox{.}(2025b)]%
        {xu2025demo}
\bibfield{author}{\bibinfo{person}{Jiawei Xu}, \bibinfo{person}{Zhandos Sembay}, \bibinfo{person}{Swathi Thaker}, \bibinfo{person}{Pamela Payne-Foster}, \bibinfo{person}{Jake~Yue Chen}, {and} \bibinfo{person}{Ying Ding}.} \bibinfo{year}{2025}\natexlab{b}.
\newblock \showarticletitle{Demo: Interactive Visualization of Semantic Relationships in a Biomedical Project's Talent Knowledge Graph}.
\newblock \bibinfo{journal}{\emph{CoRR}}  \bibinfo{volume}{abs/2501.09909} (\bibinfo{year}{2025}).
\newblock
\showeprint[arxiv]{2501.09909}~[cs.SI]
\href{https://doi.org/10.48550/arXiv.2501.09909}{doi:\nolinkurl{10.48550/arXiv.2501.09909}}


\bibitem[Xu et~al\mbox{.}(2023)]%
        {Xu2023SeqCare}
\bibfield{author}{\bibinfo{person}{Yongxin Xu}, \bibinfo{person}{Xu Chu}, \bibinfo{person}{Kai Yang}, \bibinfo{person}{Zhiyuan Wang}, \bibinfo{person}{Peinie Zou}, \bibinfo{person}{Hongxin Ding}, \bibinfo{person}{Junfeng Zhao}, \bibinfo{person}{Yasha Wang}, {and} \bibinfo{person}{Bing Xie}.} \bibinfo{year}{2023}\natexlab{}.
\newblock \showarticletitle{SeqCare: Sequential Training with External Medical Knowledge Graph for Diagnosis Prediction in Healthcare Data}. In \bibinfo{booktitle}{\emph{Proceedings of the ACM Web Conference 2023}}. \bibinfo{publisher}{ACM}, \bibinfo{address}{New York, NY, USA}, \bibinfo{pages}{2819--2830}.
\newblock
\href{https://doi.org/10.1145/3543507.3583543}{doi:\nolinkurl{10.1145/3543507.3583543}}


\bibitem[Xu et~al\mbox{.}(2025c)]%
        {xu2025fast}
\bibfield{author}{\bibinfo{person}{Yuanyuan Xu}, \bibinfo{person}{Wenjie Zhang}, \bibinfo{person}{Ying Zhang}, \bibinfo{person}{Xiwei Xu}, {and} \bibinfo{person}{Xuemin Lin}.} \bibinfo{year}{2025}\natexlab{c}.
\newblock \showarticletitle{Fast and accurate temporal hypergraph representation for hyperedge prediction}. In \bibinfo{booktitle}{\emph{Proceedings of the 31st ACM SIGKDD Conference on Knowledge Discovery and Data Mining V. 1}}. \bibinfo{pages}{1727--1738}.
\newblock


\bibitem[Yan et~al\mbox{.}(2025)]%
        {yan2024knownet}
\bibfield{author}{\bibinfo{person}{Youfu Yan}, \bibinfo{person}{Yu Hou}, \bibinfo{person}{Yongkang Xiao}, \bibinfo{person}{Rui Zhang}, {and} \bibinfo{person}{Qianwen Wang}.} \bibinfo{year}{2025}\natexlab{}.
\newblock \showarticletitle{KNOWNET: Guided health information seeking from LLMs via knowledge graph integration}.
\newblock \bibinfo{journal}{\emph{IEEE Transactions on Visualization and Computer Graphics}} \bibinfo{volume}{31}, \bibinfo{number}{1} (\bibinfo{date}{Jan} \bibinfo{year}{2025}), \bibinfo{pages}{547--557}.
\newblock
\href{https://doi.org/10.1109/TVCG.2024.3456364}{doi:\nolinkurl{10.1109/TVCG.2024.3456364}}


\bibitem[Yang and Lv(2023)]%
        {yang2023visualization}
\bibfield{author}{\bibinfo{person}{Wenzhou Yang} {and} \bibinfo{person}{Wenzhe Lv}.} \bibinfo{year}{2023}\natexlab{}.
\newblock \showarticletitle{Visualization Over Large Language Model Using Knowledge Graph}. In \bibinfo{booktitle}{\emph{2023 International Conference on Information Processing and Network Provisioning (ICIPNP)}}. \bibinfo{publisher}{IEEE}, \bibinfo{address}{Beijing, China}, \bibinfo{pages}{429--434}.
\newblock
\href{https://doi.org/10.1109/ICIPNP62754.2023.00095}{doi:\nolinkurl{10.1109/ICIPNP62754.2023.00095}}


\bibitem[You and Team(2025)]%
        {vue3}
\bibfield{author}{\bibinfo{person}{Evan You} {and} \bibinfo{person}{The Vue.js~Core Team}.} \bibinfo{year}{2025}\natexlab{}.
\newblock \bibinfo{title}{Vue.js 3 --- The Progressive JavaScript Framework}.
\newblock \bibinfo{howpublished}{\url{https://vuejs.org/}}.
\newblock
\newblock
\shownote{Vue 3 official documentation, accessed 2025-12-04}.


\bibitem[Zhang et~al\mbox{.}(2020)]%
        {zhang2020_lab_results}
\bibfield{author}{\bibinfo{person}{Zhan Zhang}, \bibinfo{person}{Daniel Citardi}, \bibinfo{person}{Aiwen Xing}, {et~al\mbox{.}}} \bibinfo{year}{2020}\natexlab{}.
\newblock \showarticletitle{Patient Challenges and Needs in Comprehending Laboratory Test Results: Mixed Methods Study}.
\newblock \bibinfo{journal}{\emph{Journal of Medical Internet Research}} \bibinfo{volume}{22}, \bibinfo{number}{12} (\bibinfo{year}{2020}), \bibinfo{pages}{e18725}.
\newblock
\href{https://doi.org/10.2196/18725}{doi:\nolinkurl{10.2196/18725}}


\bibitem[Zhao et~al\mbox{.}(2024)]%
        {zhao2024explainability}
\bibfield{author}{\bibinfo{person}{Haiyan Zhao}, \bibinfo{person}{Hanjie Chen}, \bibinfo{person}{Fan Yang}, \bibinfo{person}{Ninghao Liu}, \bibinfo{person}{Huiqi Deng}, \bibinfo{person}{Hengyi Cai}, \bibinfo{person}{Shuaiqiang Wang}, \bibinfo{person}{Dawei Yin}, {and} \bibinfo{person}{Mengnan Du}.} \bibinfo{year}{2024}\natexlab{}.
\newblock \showarticletitle{Explainability for large language models: A survey}.
\newblock \bibinfo{journal}{\emph{ACM Transactions on Intelligent Systems and Technology}} \bibinfo{volume}{15}, \bibinfo{number}{2} (\bibinfo{year}{2024}), \bibinfo{pages}{1--38}.
\newblock
\href{https://doi.org/10.1145/3639372}{doi:\nolinkurl{10.1145/3639372}}


\bibitem[Zhao et~al\mbox{.}(2025)]%
        {zhao2025medrag}
\bibfield{author}{\bibinfo{person}{Xuejiao Zhao}, \bibinfo{person}{Siyan Liu}, \bibinfo{person}{Su-Yin Yang}, {and} \bibinfo{person}{Chunyan Miao}.} \bibinfo{year}{2025}\natexlab{}.
\newblock \showarticletitle{Medrag: Enhancing retrieval-augmented generation with knowledge graph-elicited reasoning for healthcare copilot}. In \bibinfo{booktitle}{\emph{Proceedings of the ACM Web Conference 2025}}. \bibinfo{publisher}{Association for Computing Machinery}, \bibinfo{address}{New York, NY, USA}, \bibinfo{pages}{4442--4457}.
\newblock
\href{https://doi.org/10.1145/3696410.3714782}{doi:\nolinkurl{10.1145/3696410.3714782}}


\end{thebibliography}
\appendix

\end{document}